\documentclass{JHEP}
\usepackage{epsfig}

\def\eq#1{{eq.~(\ref{#1})}}
\def\eqs#1#2{{eqs.~(\ref{#1})--(\ref{#2})}}

\let\vev\VEV

\def\Im{\mathop{\mbox{Im}}}
\def\Re{\mathop{\mbox{Re}}}
\def\Tr{\mathop{\mbox{Tr}}\,}

\def\qq{$\vev{\bar q q}$ }
\def\GG{$\vev{\alpha_s GG/ \pi}$ }

\def\ee{$\varepsilon'/\varepsilon$ }
\newcommand{\bea}{\begin{eqnarray}}
\newcommand{\beq}{\begin{equation}}
\newcommand{\eea}{\end{eqnarray}}
\newcommand{\eeq}{\end{equation}}
\newcommand{\nnu}{\nonumber}

\title{The $\Delta I = 1/2$ Rule and $\widehat{B}_{K}$\\ 
 at  $O(p^4)$ in the Chiral Expansion.}

\author{S. Bertolini$^{\dag \S}$, J.O. Eeg$^{\ddag}$,
M. Fabbrichesi$^{\dag \S}$ and E.I. Lashin$^{\dag}$\thanks{Permanent address:
Ain Shams University, Faculty of Science, Dept. of Physics, Cairo, Egypt.}\\
$^{\dag}$ INFN, Sezione di Trieste\\
$^{\S}$ Scuola Internazionale Superiore di Studi Avanzati\\
via Beirut 4, I-34013 Trieste, Italy.\\
$^{\ddag}$ Fysisk Institutt, Universitetet i Oslo\\
N-0316 Oslo, Norway.}

\abstract{We calculate the
hadronic matrix elements to $O(p^4)$ in the chiral expansion for
the ($\Delta S =1$) $K^0 \rightarrow 2\, \pi$ decays and the 
($\Delta S=2$) $\bar K^0$-$K^0$ oscillation. This is done within the framework
of the chiral quark model.
The chiral coefficients thus determined depend on the values of the quark
and gluon condensates and the constituent quark mass.
We show that it is possible  to fit
the $\Delta I =1/2$ rule of kaon decays with values of
the condensates close to those obtained by QCD sum rules.
The renormalization invariant
amplitudes are obtained by matching the hadronic matrix elements and their
chiral corrections to the short-distance NLO Wilson coefficients.
For the same input values, we study the parameter $\widehat B_K$ of kaon
oscillation and find $\widehat B_K = 1.1 \pm 0.2$. 
As an independent check, we determine $\widehat B_K$ from the 
experimental value of the $K_L$-$K_S$ mass difference
by using our calculation of the long-distance contributions.
The destructive interplay between the short- and long-distance 
amplitudes yields $\widehat B_K = 1.2 \pm 0.1$, in agreement 
with the direct calculation.}

\preprint{SISSA 4/97/EP \\
April 1997 \\
Revised, September 1997}

\begin{document}

\section {Introduction}	

The physics of kaons is an important testing ground for our understanding of
low-energy QCD. The hadronic matrix elements for the nonleptonic decays
must satisfy the non-trivial constraint of the $\Delta I =1/2$ rule
that shows the striking enhancement of the decay width in which the two final
pions combine in the isospin $I=0$ state with respect to the  
$I=2$ state. The interplay between the renormalization-group evolution of
the Wilson coefficients and the matrix elements of the relevant quark 
operators in both $\bar K^0$-$K^0$
mixing and $K^0 \rightarrow 2 \, \pi$ decays determines the
theoretical predictions of the
$CP$-violating parameters $\varepsilon$ and $\varepsilon'/\varepsilon$,
the knowledge of which is essential to the standard model.

The recent progress in the next-to-leading-order (NLO) computation of the
Wilson coefficients of the $\Delta S =1$~\cite{NLO1} and 
$\Delta S =2$~\cite{NLO2} effective
lagrangian at the quark level makes more urgent the need to bring
under better control
the non-perturbative estimate of the corresponding hadronic matrix elements.
Different approaches have been pursued, ranging from lattice 
simulations~\cite{reticolo}, $1/N_c$-expansion~\cite{BBG}, phenomenological
approaches~\cite{Buras}
dispersion-integral 
techniques~\cite{Pich} and low-energy QCD modeling~\cite{PdeR,BBdeR}. 
A particular
case of the latter, the chiral quark model ($\chi$QM)~\cite{QM,QM1} 
has been analyzed in
detail in a series of papers~\cite{I,II,III} with the encouraging result of
fitting the $\Delta I =1/2$ rule by a consistent and 
reasonable choice of the input parameters.
These are essentially three:  the quark condensate, the
gluon condensate and   $M$---a parameter of the model which
corresponds to a typical constituent quark mass.

In a previous paper~\cite{I}, we have calculated
the coefficients of all the $O(p^2)$ terms of the $\Delta S = 1$
chiral lagrangian and included the corresponding chiral loop
renormalizations (which are of $O(p^4)$).
The present paper completes our previous analyses 
\begin{itemize}
\item
by including in the hadronic matrix elements the complete NLO
$O(p^4)$ corrections;
\item
by updating the short-distance calculation of the Wilson coefficients
according to the most recent determinations of $\alpha_s$ and $m_t$.
\end{itemize}

The determination of
the $K \rightarrow 2\pi$
matrix elements at $O(p^4)$ is a
non-trivial computation. To our knowledge it is the
first time that the whole of these 
hadronic matrix elements are estimated to this order
by any technique at all (for previous discussion, 
see refs.~\cite{bef0,meson}).

The introduction of these corrections allows us
to determine a consistent range of values of
the input parameters for which the $\Delta I=1/2$ rule is 
reproduced with an accuracy at the 20\% level.  
We find the best fit to this
order in the chiral expansion
for values of the condensates close to those
derived by QCD sum rules and a value of $M$ in agreement with independent
estimates based on radiative kaon decays~\cite{Bijnens}.
A coherent picture is thus provided: all hadronic matrix elements
are computable for a common set of input parameters and no ad-hoc assumption
is necessary to fit the $\Delta I =1/2$ rule. 

The hadronic matrix elements thus found are eventually
matched together with one-loop
chiral corrections to the corresponding NLO Wilson coefficients. The matching
scale is a delicate choice in so far as it can neither be too high, in order
for chiral perturbation theory to remain valid, nor too low, for the
Wilson coefficients to be reliable. We choose a scale of 0.8 GeV that we
identify with the chiral symmetry breaking scale. 
Even though at this scale $\alpha_s$ is already
rather large, we have verified that the computed
observables change by no more than 30\% when moving from LO to NLO order
Wilson coefficients.  
By considering the shift in the value of $\alpha_s$ at $\mu=0.8$ GeV 
when going from the NLO to the next order, we can estimate that the neglect
of next-to-NLO corrections in the running of the Wilson coefficients
may affect our predictions at the 10\% level,
which is within the ``systematic'' error we assign to our analysis.

Once the input parameters of the model have been fixed,
other quantities of kaon physics can also be
estimated. In particular we discuss the deviation from the vacuum
insertion approximation (VSA) in $\bar K^0$-$K^0$
oscillations that is parameterized by the scale independent parameter
$\widehat B_K$. 
The inclusion of
the $O(p^4)$ corrections to the $\Delta S=2$ chiral lagrangian
provides us---for the input parameters
identified by the $\Delta I =1/2$ selection rule---with a central value of
 $\widehat B_K \simeq 1.1$. 

Contrarily to what one might
think, such a large value is not
in conflict with the experimental
determination
of the $K_L$-$K_S$ mass difference $\Delta M_{LS}$.
To prove this point,
we have computed the long-distance mesonic contributions that arise
from the double insertion of the $\Delta S=1$ chiral lagrangian,
and find that they are about 20--30\% of the whole mass difference and of the
opposite sign with respect to the short-distance part~\cite{IV}.
By requiring 
that the $\Delta M_{LS}$ thus calculated
fits the experimental value, we obtain  
$\widehat B_K \simeq 1.2$ in striking agreement with the direct
calculation.

We thus present two independent estimates of $\widehat B_K$
within the $\chi$QM approach:
\begin{itemize}
\item
$\widehat B_K = 1.1 \pm 0.2$ from direct calculation,
\item
$\widehat B_K = 1.2 \pm 0.1$ from $\Delta M_{LS}$,
including long-distance contributions,
\end{itemize}
where the errors include a flat variation of all relevant input
parameters.
According to these results,
values of $\widehat B_K$ smaller than one
 are disfavored in the $\chi$QM. 

These results can then be applied to the prediction of the direct
$CP$ violating parameter \ee .
Because of the importance and the complexity of such a calculation,
we have decided to present it in an independent paper~\cite{VI}. 

In order for the present paper to be as self-contained as possible,
we have included
in the following two subsections the relevant lagrangians and a brief
introduction to the $\chi$QM. Such  introductions summarize those of
our previous papers, to which we refer the reader for further
details and references. 

\subsection{The Quark Effective Lagrangian}

Let us introduce our notation by recalling that the $\Delta S=1$
quark effective lagrangian at a scale $\mu < m_c$ can be written
as~\cite{GW}
 \beq
{\cal L}_{\Delta S = 1} = - C_i(\mu)\ Q_i (\mu) =
\frac{G_F}{\sqrt{2}} V_{ud}\,V^*_{us} \sum_i \Bigl[
z_i(\mu) + \tau y_i(\mu) \Bigr] Q_i (\mu) 
 \, . \label{Lquark}
\eeq

The $Q_i$ are effective 
four-quark operators obtained by integrating out in the standard
model the vector bosons and the heavy quarks $t,\,b$ and $c$. 
A convenient and by now standard
basis includes the following twelve quark operators:
 \beq
\begin{array}{rcl}
Q_{1} & = & \left( \overline{s}_{\alpha} u_{\beta}  \right)_{\rm V-A}
            \left( \overline{u}_{\beta}  d_{\alpha} \right)_{\rm V-A}
\, , \\[1ex]
Q_{2} & = & \left( \overline{s} u \right)_{\rm V-A}
            \left( \overline{u} d \right)_{\rm V-A}
\, , \\[1ex]
Q_{3,5} & = & \left( \overline{s} d \right)_{\rm V-A}
   \sum_{q} \left( \overline{q} q \right)_{\rm V\mp A}
\, , \\[1ex]
Q_{4,6} & = & \left( \overline{s}_{\alpha} d_{\beta}  \right)_{\rm V-A}
   \sum_{q} ( \overline{q}_{\beta}  q_{\alpha} )_{\rm V\mp A}
\, , \\[1ex]
Q_{7,9} & = & \frac{3}{2} \left( \overline{s} d \right)_{\rm V-A}
         \sum_{q} \hat{e}_q \left( \overline{q} q \right)_{\rm V\pm A}
\, , \\[1ex]
Q_{8,10} & = & \frac{3}{2} \left( \overline{s}_{\alpha} 
                                                 d_{\beta} \right)_{\rm V-A}
     \sum_{q} \hat{e}_q ( \overline{q}_{\beta}  q_{\alpha})_{\rm V\pm A}
\, , \\[1ex]
Q_{11} & = & \frac{g_s}{16 \pi^2} \: \bar s \: \left[ 
m_d \left(1 + \gamma_5 \right) + m_s \left(1 - \gamma_5 \right) \right] 
\: \sigma \cdot G \: d
\, , \\[1ex]
Q_{12} & = & \frac{e}{16 \pi^2} \: \bar s \: \left[ 
m_d \left(1 + \gamma_5 \right) + m_s \left(1 - \gamma_5 \right) \right] 
\: \sigma \cdot F \: d
\, , 
\end{array}  
\label{Q1-10} 
\eeq
where $\alpha$, $\beta$ denote color indices ($\alpha,\beta
=1,\ldots,N_c$) and $\hat{e}_q$  are quark charges. Color
indices for the color singlet operators are omitted. 
The subscripts $(V\pm A)$ refer to
$\gamma_{\mu} (1 \pm \gamma_5)$.
We recall that
$Q_{1,2}$ stand for the $W$-induced current--current
operators, $Q_{3-6}$ for the
QCD penguin operators and $Q_{7-10}$ for the electroweak penguin (and box)
ones. The quark operators $Q_{11,12}$, involving the gluon and photon
 fields, are the dipole penguin 
operators whose $K^0\to\pi\pi$ matrix elements
arise at $O(p^4)$~\cite{bef0}.

Even though not all the operators in \eq{Q1-10} are independent, this basis
is of particular interest for 
the present numerical analysis because it is that employed
for the calculation of the Wilson coefficients 
to the NLO in $\alpha_s$~\cite{NLO1}.

In the present paper we will mainly discuss the features related
to the first six 
operators in (\ref{Q1-10}) because the electroweak penguins
$Q_{7-10}$ have their contributions
suppressed by the smallness of their $CP$ conserving Wilson coefficients, 
while $Q_{11-12}$ have very small matrix elements~\cite{bef0}. 
The electroweak operators play however a crucial role
in the discussion of \ee~\cite{VI}. 

The functions $z_i(\mu)$ and $y_i(\mu)$ are the
 Wilson coefficients and $V_{ij}$ the
Koba\-ya\-shi-Mas\-kawa (KM) matrix elements; $\tau = - V_{td}
V_{ts}^{*}/V_{ud} 
V_{us}^{*}$. 
The numerical values of the Wilson coefficients at a given scale
depend on $\alpha_s$. We take the most recent world average~\cite{alfa}
\beq
\alpha_s (m_Z) = 0.1189 \pm 0.0020 \, ,
\eeq
which at the NLO corresponds approximately to
\beq
\Lambda^{(4)}_{\rm QCD} = 340 \pm 40 \: \mbox{MeV} \, .
\label{lambdone}
\eeq
We match the Wilson coefficients at the $m_W$ scale with the full
electroweak theory by using the LO $\overline{MS}$ running
top mass $m_t(m_W)=177\pm 7$ GeV which
corresponds to the pole mass $m_t^{pole}=175\pm 6$ GeV~\cite{mt}. 
For the remaining quark
thresholds we take $m_b(m_b) =4.4$ GeV and $m_c(m_c)=1.4$ GeV.

Similarly, the  effective $\Delta S=2$ quark 
lagrangian at scales $\mu<m_c$ can be written as
\bea
{\cal L}_{\Delta S = 2} 
&=& -C_{S2}(\mu)\ Q_{S2}(\mu) \nnu \\
&=& -\frac{G_F^2 m_W^2 }{4 \pi^2} \left[\lambda_c^2 \eta_{1} S(x_c) 
+ \lambda_t^2 \eta_{2} S(x_t)
 + 2 \lambda_c \lambda_t \eta_3 S(x_c , x_t)\right] b(\mu) Q_{S2}(\mu)
\label{lags2}
\eea
where
$\lambda_j = V_{jd} V_{js}^{*}$ and
$x_i = m_i^2 / m_W^2$.
We denote by
$Q_{S2}$ the $\Delta S=2$ local four quark operator
\beq
Q_{S2} =(\bar{s}_L \gamma^{\mu} d_L) (\bar{s}_L \gamma_{\mu} d_L)
\, , \label{QS2}
\eeq
which is the only numerically relevant operator in this case.

The integration of the electroweak loops leads to the
functions $S(x)$ and $S(x_c, x_t)$, the exact
expressions of which can be found in ref.~\cite{Inami-Lim}.
They describe the dependence of the $\Delta S = 2$ transition amplitude
on the masses of the charm and top quarks 
in the absence of strong interactions.

The short-distance QCD corrections are encoded in the coefficients $\eta_1$,
$\eta_2$ and $\eta_3$ and $b(\mu)$. The $\eta_i$ coefficients represent
the renormalization effects down to the scale $m_c$. 
They are functions of the heavy quarks masses and of
$\Lambda_{\rm QCD}$.
These QCD corrections are available to NLO in the strong
coupling~\cite{NLO2}.
The scale-dependent function $b(\mu)$ describes the overall running
below the charm threshold and it is given by
\beq
b ( \mu ) = \left[\alpha_s\left(\mu\right)\right]^{-2/9}
\left( 1 - J_3 \frac{\alpha_s\left(\mu\right)}{4 \pi} \right) \label{b}
\, ,
\label{wc}
\eeq
where $J_3$ depends on the $\gamma_5$-scheme used in the regularization. The
naive dimensional regularization (NDR) and 
the 't Hooft-Veltman (HV) scheme yield, respectively:
\beq
J_3^{\rm NDR} = -\frac{307}{162} \quad \quad \mbox{and} \quad \quad
 J_3^{\rm HV} =-\frac{91}{162} \, .
 \eeq

\subsection{The Chiral Quark Model}

In order to evaluate the bosonization
of the quark operators in eqs. (\ref{Q1-10}) and (\ref{QS2})
we exploit  the $\chi$QM approach
which provides an effective link between  QCD and
chiral perturbation theory.

The $\chi$QM  can be thought of as
the mean field approximation of the extended Nambu-Jona-Lasinio
(ENJL) model for low-energy
QCD. A detailed discussion of the ENJL model and
its relationship with QCD---as well as with the $\chi$QM---
can be found, for instance, in ref.~\cite{BBdeR}. 

In the $\chi$QM, the light (constituent) 
quarks are coupled to the Goldstone mesons by  
the term
\beq
 {\cal{L}}^{\rm int}_{\chi \mbox{\scriptsize QM}} = 
- M \left( \overline{q}_R \; \Sigma q_L +
\overline{q}_L \; \Sigma^{\dagger} q_R \right) \, ,
\label{M-lag}
\eeq
where $q^T\equiv (u,d,s)$ is the quark flavor triplet, and
the $3\times 3$ matrix
\beq
\Sigma \equiv \exp \left( \frac{2i}{f} \,\Pi (x)  \right)
\label{sigma}
\eeq
contains the pseudoscalar octet
 $\Pi (x) = \sum_a \lambda^a \pi^a (x) /2 $,
$(a=1,...,8)$. The scale
$f$ is identified at the tree level with the  pion decay constant $f_\pi$ 
(and equal to $f_K$ before chiral loops and higher order
corrections are introduced). In the $\chi QM$,
the pion decay constant 
is given by a logarithmic divergent quark loop integral $f_\pi^{(0)}$,
which is numerically identified with $f$, such that
$f_+ \equiv f_\pi^{(0)}/f = 1$.

The $\chi$QM has been discussed
in several works over the years~\cite{QM,QM1}. 
We opted for the somewhat more restrictive definition
suggested in ref.~\cite{QM1} 
(and there referred to as the QCD effective action
model) in which the meson degrees of freedom
do not propagate in the original lagrangian.

The QCD gluonic fields are considered as
integrated out down to the  chiral breaking  scale $\Lambda_\chi$, here  acting
as an infrared cut-off. The effect of the 
remaining low-frequency modes are assumed to be
well-represented by gluonic vacuum condensates, the leading contribution
coming from 
\beq
\langle \frac{\alpha_s}{\pi} G G \rangle \, .
\eeq
The constituent quarks are taken to be
propagating in the fixed background of the soft gluons.
This defines an effective QCD lagrangian 
$ {\cal{L}}^{\rm eff}_{\rm QCD} (\Lambda_\chi) $,
whose propagating fields are the $u,d,s$ quarks.
The strong $\chi$QM lagrangian is therefore given by
\beq
{\cal{L}}_{\chi \mbox{\scriptsize QM}} =
{\cal{L}}^{\rm eff}_{\rm QCD} (\Lambda_\chi) +
 {\cal{L}}^{\rm int}_{\chi \mbox{\scriptsize QM}}\ .
\label{LchiQM}
\eeq

The ${\cal{L}}_{\chi \mbox{\scriptsize QM}}$ interpolates between 
the chiral breaking scale
$\Lambda_{\chi}$ and $M$ (the constituent quark mass).
The three light quarks ($u,\ d,\ s$) are the only dynamical degrees of freedom
present within this range. 
The total quark masses are given by $M+m_q$ where
$m_q$ is the current quark mass in the QCD lagrangian.
The Goldstone bosons and
the soft QCD gluons are taken in our approach as external fields.
A kinetic term for the mesons, as well as the complete
chiral lagrangian,
is generated and determined by
integrating out the constituent quark degrees of freedom of the model.
By combining \eq{LchiQM} with eqs. (\ref{Lquark}) and (\ref{QS2})
one may obtain the $\Delta S =1$ and $\Delta S=2$ weak
chiral lagrangians as effective theories of
the $\chi$QM.
In the matching process, the many coefficients of the chiral
lagrangian are determined---to the order $O(\alpha_s N_c)$ in our
computation---in
terms of $M$, the quark and gluon condensates.
We neglect heavier scalar, vector and axial meson multiplets.

In conventional chiral perturbation theory the scale dependence of  
meson loops renormalization is canceled by construction by the 
$O(p^4)$ counterterms in the chiral lagrangian.
While in our approach this is mantained for the strong sector
of the chiral lagrangian, the tree-level 
counterterms of the weak sector are taken to be $\mu$ 
independent and a scale dependence is introduced in the hadronic
matrix elements via the meson 
loops, evaluated in dimensional regularization
with minimal subtraction. This scale dependence
is eventually matched with that of the Wilson coefficients. 
Tree level counterterms acquire a scale dependence
at the next order in the chiral expansion, via meson loop
renormalization.

A more detailed discussion about our approach including the complete
determination of the $\Delta S = 1$ chiral lagrangian at $O(p^2)$
can be found in refs.~\cite{I,bef0}.

\section{Building Blocks}

Within the $\chi$QM, matrix elements of the quark operators $Q_{1-10}$
can be calculated in the factorizable approximation as products
of two currents or two densities. Non-factorizable
matrix-elements are proportional to the gluon condensate. 
In this way  the hadronic matrix elements of the relevant operators are
constructed in terms of products of two building blocks.
These elementary blocks must be computed to the 
appropriated order in
$p^2$ and $m_q$ to yield the $O(p^4)$ matrix elements. 

By integrating over quark loops we obtain
the following  matrix elements of quark densities,
based on the  lagrangian in \eq{LchiQM}:
 \bea
 \langle0|\,\overline{s} \gamma_5 u\, |K^+(k)\rangle 
&=&  i \sqrt{2} \;
\left\{ \frac{\vev{\overline{q} q}}{f} - k^2 \,
\frac{f}{2 M} \left( f_+ + \frac{k^2}{\Lambda_\chi^2} \right) \right.  
\nnu \\
&& + \left. \left( m_s + m_u \right)  \: f \left(
 f_+ + 3 \, \frac{k^2}{\Lambda_\chi^2} \right)\right. \nnu \\
&& + \left. \frac{f}{M}\left[(m_s^2 + m_u^2)
\left(f_+ - 6\frac{M^2}{\Lambda_\chi^2}\right)  
- m_s m_u\ f_+ \right] \right\} \; ,
\label{Kvacuum}\\
\langle \pi^+(p_+)|\,\overline{s} d\, |K^+(k)
\rangle \, 
&=&
\, - \frac{\langle \overline{q} q \rangle}{f^2}
\, + \frac{q^2}{2 M} f_+ 
 +  \frac{3 M}{2 \Lambda_\chi^2}\Big( P^2 - q^2 \Big) \nnu \\
&& + \frac{1}{16 M \Lambda_\chi^2} \Big( P^4 + 2 P^2 q^2 + 5 q^4 \Big) \nnu \\
&& - \left( m_s + m_d \right) \left[ f_+ + 2 \, \frac{q^2}{\Lambda_\chi^2}
\right] - \left( m_s - m_d \right)  \, \frac{q\cdot P}{\Lambda_\chi^2} \nnu \\
&& - 2 \, m_u\left[ f_+ + \frac{3 P^2 + 5 q^2}{4\Lambda_\chi^2} \right]\nnu \\
&& - \frac{1}{M}\left[(m_s^2 + m_s m_d + m_d^2)
\left(f_+ - 6\frac{M^2}{\Lambda_\chi^2}\right) \right. \nnu \\  
&& \hspace*{1.3cm} 
\left. -\ 6\ m_u (m_s  + m_d  + m_u )\frac{M^2}{\Lambda_\chi^2}
\right] \, ,
\label{Kpi}
\eea
where $\langle \overline{q} q \rangle$ is the quark condensate for
zero current quark mass.
In the calculation of the four-fermion matrix elements at $O(p^4)$
also the density amplitude
$\langle\pi\pi|\,\overline{s}\gamma_5 d\,|K^0\rangle$ contributes.
A discussion on its evaluation is given in appendix A.

The corresponding quark current matrix elements are given by:
\bea
\langle 0|\,\overline{s} \gamma^\mu \left(1 - \gamma_5\right) u\,|K^+(k)
\rangle 
&=& - i \sqrt{2}\ f\ \left\{
 f_+ + \frac{k^2}{\Lambda_\chi^2} \right. \nnu \\
&& + \left. \frac{m_s + m_u}{2M} \left[ f_+
- 12 \frac{M^2}{\Lambda_\chi^2} \right] \right\} k^\mu 
\label{fpi}\\
\langle\pi^+(p_+)|\,\overline{s}\gamma^\mu\left(1-\gamma_5\right) d\,|K^+(k)
\rangle \, 
&=& - \left[
f_+ + \frac{P^2 + 3 q^2}{2\Lambda_\chi^2} \right] P^\mu + 
\frac{q \cdot P}{\Lambda_\chi^2} q^\mu \nnu \\
&& + \frac{3M ( m_s + m_d + 2 m_u)}{\Lambda_\chi^2} P_\mu \nnu \\
&& - \frac{m_s - m_d}{2M} \left[ f_+ - 6 \frac{M^2}{\Lambda_\chi^2} 
\right]q^\mu  \; ,
\label{f+}\\
\langle 0|\,\overline{s} \gamma^\mu T^a\left(1 - \gamma_5\right) u\,|K^+(k)
\rangle 
&=& - \frac{i g_s \sqrt{2}}{16 \pi^2 f} \, G_{\nu\tau}^a A_G^{\mu\nu\tau}(k)
 \label{fpiglue} \\
\langle\pi^+(p_+)|\,\overline{s}\gamma^\mu T^a\left(1-\gamma_5\right) d\,
|K^+(k) \rangle \, 
&=& -  \frac{g_s}{16 \pi^2 f^2} \, G_{\nu\tau}^a B_G^{\mu\nu\tau} (q,P)
\label{f+glue}
\eea
where  $q=k-p_+$ and $P=k+p_+$, $k$ being the incoming momentum of the
kaon field and $p_+$ the outgoing momentum of the pion field.
The chiral symmetry breaking scale is taken as $\Lambda_\chi =
2\pi\sqrt{6/N_c} f \simeq 0.8$ GeV,
while $f_+ \equiv f^{(0)}_\pi/f = 1$ is the vector form factor at
zero momentum transfer. Finally, $G_{\nu\tau}^a$ with $a=1,...,8$
is the usual $SU(3)_c$ gluon field tensor and $T^a$ are the $SU(3)_c$ 
generators normalized as $\Tr T^aT^b = \delta^{ab}/2$.

The \eqs{fpiglue}{f+glue} represent the gluonic
corrections, which are computed by using one-gluon dressed quark
propagators and color Fierz transformations
on the four-quark operators
(the $SU(3)_c$ index $a$ is to be summed over in the full matrix
element). They contribute to the non-factorizable part of the hadronic
matrix elements via the relation
\beq
 g_s^2  G^a_{\mu\nu}G^a_{\alpha\beta} =
\frac{\pi^2}{3}\langle \frac{\alpha_s}{\pi}GG \rangle
\left(\delta_{\mu\alpha}\delta_{\nu\beta} -
\delta_{\mu\beta}\delta_{\nu\alpha}\right)
\label{gluonaverage}
\eeq
which allows us to express our results
in terms of the gluonic vacuum condensate.

The two gluonic form factors are given by:
\bea
A_G^{\mu\nu\tau}(k) & =& \epsilon (k,\mu,\nu,
\tau) \left[ 1 + \frac{k^2}{6 M^2} - \frac{m_s + m_u}{2 M} \right]
\nnu \\
&&   +\ i 
\frac{m_s - m_u}{4 M} \left[ k_\nu g_{\mu\tau} - k_\tau g_{\mu\nu} \right]
\eea
and
\bea
B_G^{\mu\nu\tau} (q,P) & = &  \epsilon (P,\mu,\nu,
\tau) \left[ 1 + \frac{P^2}{12 M^2} + \frac{q^2}{4M^2} \right]  \nnu \\
&& - \epsilon (q,\mu,\nu,\tau) \frac{q \cdot P}{6 M^2}
-  \epsilon (P,q,\mu,\nu) \frac{q_\tau}{12 M^2} +  \epsilon (P,q,\mu,\tau)
\frac{q_\nu}{12 M^2} \nnu \\
&& - i \frac{2}{3} \Big( q_\nu g_{\mu\tau} - q_\tau g_{\mu\nu} \Big)
- i \frac{P_\mu}{30 M^2}   \Big( P_\nu q_\tau - P_\tau q_\nu \Big) \nnu \\
&& + i \frac{7 q \cdot P}{120 M^2}  
\Big( P_\nu g_{\mu\tau} - P_\tau g_{\mu\nu} \Big) 
- i \left[ \frac{29 P^2}{240 M^2} +
\frac{11 q^2}{80 M^2} \right]
  \Big( q_\nu g_{\mu\tau} - q_\tau g_{\mu\nu} \Big) \nnu \\
& & - \frac{m_s}{12 M} \left\{ 
\epsilon\left[ (5 P + q), \mu ,\nu ,\tau \right]
+ \frac{i}{4} \left[ \left(P_\nu - 19 q_\nu \right) g_{\mu\tau} - \left(\nu
\rightarrow \tau \right)\right] \right\} \nnu \\
& &  - \frac{m_d}{12 M} \left\{ 
\epsilon\left[ (5 P - q), \mu ,\nu ,\tau \right]
- \frac{i}{4} \left[ \left(P_\nu + 19 q_\nu \right) g_{\mu\tau} - \left(\nu
\rightarrow \tau \right)\right]\right\}  \nnu \\
& &  - \frac{m_u}{12 M} \left\{12  \epsilon \left(  P, \mu ,\nu ,\tau \right)
- \frac{i}{4} \left[ 14 q_\nu g_{\mu\tau} - \left(\nu
\rightarrow \tau \right)\right] \right\}   
\eea
where $\epsilon (k, \mu ,\nu ,\tau) \equiv 
k^\alpha\ \epsilon_{\alpha\mu\nu\tau}$.

The procedure by which it is possible to determine the hadronic matrix elements
and the corresponding chiral 
coefficient to the desired order
for all the relevant  operators by means
of the given building blocks is discussed in refs.~\cite{I,bef0},
to which we refer the interested reader.

Before proceeding it is important to
identify what parts of the building blocks are absorbed by the 
renormalization of the parameters and fields
of the lagrangian. To this we now turn. 

\section {Wave-Function and Coupling-Constant Renormalizations}

To the order $O(p^4)$ at which we are working, both 
the wave-function and the meson
decay constant renormalizations must be included. They give sizeable
contributions that cannot be neglected. In our computation,
the renormalization comes in two parts which
are conceptually
distinct. On the one hand we have the chiral renormalization
generated by the meson loops, to which we have to add
the renormalization that is 
specific of the $\chi$QM and
originates from the expansion of the building blocks beyond the leading order.
In refs. \cite{I,II,III} only the chiral loop renormalizations were included.

\subsection{Wave-Function Renormalizations}

The wave-function renormalizations which arise in the $\chi$QM
from direct calculation of the  $K \rightarrow K$ and 
$\pi \rightarrow \pi$ propagators are given at $O(p^2)$ by:
\bea
Z_K &= & 1 - 2 \frac{m_K^2}{\Lambda_{\chi}^2} 
+ 6 \frac{M (m_s + m_u)}{\Lambda_{\chi}^2} \nnu \\
Z_\pi &= & 1 - 2 \frac{m_\pi^2}{\Lambda_{\chi}^2} 
+ 6 \frac{M (m_d + m_u)}{\Lambda_{\chi}^2}
\label{waves}
\eea
The complete $O(p^4)$ 
expressions are given in appendix B. 
The renormalizations above are added to those induced
by the one-loop chiral corrections.

\subsection{Renormalization of $f_K$ and $f_\pi$ in the $\chi QM$}

The building block in \eq{fpi} gives directly the 
 corrected (unrenormalized)  version of
the coupling constant $f$ yelding (for instance for $f_K$)
\beq
f_K^U = f \left[ 1 + \frac{m_K^2}{\Lambda^2_\chi} +
 \frac{m_s + m_u}{2 M}
\left( 1 - 12 \frac{M^2}{\Lambda^2_\chi} \right) \right]
\eeq
However, it is only after inclusion of the wave-function renormalization
in \eq{waves} that we find the complete $\chi$QM expression:
\beq
f_K^R = \sqrt{Z_K} f_K^U = f \left[ 1 + \frac{m_s + m_u}{2 M}
\left( 1 - 6 \frac{M^2}{\Lambda^2_\chi} \right) \right]
\label{fkQM}
\eeq
and similarly for $f_\pi$. 
Notice that $f_{\pi,K}^R$ coincide at $O(p^4)$ with the physical decay
constants $f_{\pi,K}$ only after inclusion of the proper chiral loop
renormalizations (see next subsection).

The consistency of the entire
procedure can be verified by considering the renormalization
of the charged form factor $f_+(q^2)$ in the coupling of mesons to
an external vector field (photon or vector boson). This form factor
 can be read off eq. (\ref{f+}).
After renormalization we find that the on-shell form factor is
\beq
f^R_+ (q^2) = \sqrt{Z_K}\sqrt{Z_\pi} f^U_+ (q^2) = 1 + 
\frac{q^2}{\Lambda^2_\chi} \, ,
\eeq
which correctly preserves the current-conservation condition  $f_+ (0) = 1$.

By matching 
the expressions above for the renormalized $f_{K}$ and $f_\pi$
with the chiral lagrangian results 
we obtain the $\chi$QM determinations (see also appendix B)
\bea
L_4 &=& 0\ , 
\label{L4QM} \\
L_5 &=& -  \frac{f^4}{8 M \langle \bar{q} q \rangle} \left(
           1- 6 \frac{M^2}{\Lambda_\chi^2} \right)  \, ,
\label{L5QM}
\eea 
where by PCAC
\beq
\langle \bar{q} q \rangle = \, - \, \frac{f^2 m_K^2}{m_s+m_u} \; 
            = \, - \,\frac{f^2 m_\pi^2}{m_d+m_u} \;  .
\label{Cond}
\eeq 
The above equations are obtained at the constituent quark mass scale $M$,
where the quark fields
 are integrated out and the standard chiral lagrangian arise
as the effective theory of the $\chi$QM.
The parameter $L_4$ is vanishing in the $\chi$QM
up to higher order gluon condensate contributions.
The usual numerical determinations of the renormalized
$L_4$ and $L_5$ in chiral perturbation theory must then be run down to this
scale in order to be compared to the result of the $\chi$QM,
after taking into account the different subtraction used in the 
renormalization of the chiral loops.
We address this issue in the next subsection.

\subsection{The Chiral Coupling $f$ at the One-Loop Order}

The one-loop expressions for $f_\pi$ and $f_K$ allow us
to determine the value of the chiral coupling 
$f$ to be used in the NLO calculation.
In order to compare \eq{fkQM}, and the analogous one for $f_\pi$,
with the physical values of the decay constants, 
the $O(p^4)$ renormalization induced
by chiral loops has to be included. 
In terms of the tree-level counterterms of the $O(p^4)$ strong chiral
lagrangian, one obtains the equations~\cite{2-eqs}
\beq
f^2 - f_\pi f  - 2 h(m_\pi,\mu) - h(m_K,\mu) +  
       4 m_\pi^2 L_5(\mu) + (4 m_\pi^2 + 8 m_K^2) L_4(\mu) = 0 \, ,
\label{f0p}
\eeq
for the pion decay constant $f_\pi$, and
\beq
f^2 - f_K f  - \frac{3}{4}  h(m_\pi,\mu) -  \frac{3}{2} h(m_K,\mu)
 - \frac{3}{4} h(m_\eta,\mu)
   + 4 m_K^2  L_5(\mu) + (4 m_\pi^2 + 8 m_K^2) L_4(\mu) = 0 \, ,
\label{f0k}
\eeq
for $f_K$, where, by keeping only the logarithmic part from the 
loop integrals,
\beq
h(m,\mu) = \frac{m^2}{32 \pi^2} \ln \frac{m^2}{\mu^2}
\, .
\eeq
This prescription defines the renormalized couplings $L_5$ and $L_4$.
Once $L_4$ is given, these equations
may be solved for $L_5$ and $f$.

The entire procedure contains many uncertainties, the input value of $L_4$
being one of them. According to the original derivation of ref.~\cite{2-eqs} 
we take $L_4 (m_\eta) =0.0\pm 0.5$. 
Using the known anomalous dimensions of the couplings $L_i$~\cite{2-eqs}, we
run $L_4$ up to the matching scale $\mu = \Lambda_\chi\simeq 0.8$ GeV 
where we find
\beq
L_4 (\Lambda_\chi) = - (0.3\pm 0.5) \times 10^{-3} \, .
\label{L4Lchi}
\eeq
For this value, the solution of the system  of \eq{f0p} and \eq{f0k}
at the scale $\mu =\Lambda_\chi$ gives 
\beq
L_5 (\Lambda_\chi) = (1.2 \pm 0.4) \times 10^{-3} \quad \mbox{and} \quad 
f = 0.086 \pm 0.013 \; \mbox{GeV} \, , 
\label{f0}
\eeq
where the errors come from uncertainties in the input parameters, mainly
those for $L_4$.

By solving the same equations at the scale $M\simeq 0.2$ GeV, where
\beq
L_4 (M) =  (0.8\pm 0.5) \times 10^{-3} \ ,
\label{L4M}
\eeq
we obtain
\beq
L_5 (M) = (4.3 \pm 0.4) \times 10^{-3} 
\label{f0M}
\eeq
and
$f = 0.086 \pm 0.013$ GeV, as expected from the scale-independence of $f$.

A consistent solution of \eqs{f0p}{f0k} is also obtained by using 
 the $\overline{MS}$ result for $h(m,\mu)$, that is
\beq
h(m,\mu) = \frac{m^2}{32 \pi^2} \left(\ln \frac{m^2}{\mu^2} - 1 \right)
\, .
\eeq
In this respect,
it is important to remark that
the determination of $L_5$ and $L_4$
depend on the subtraction scheme employed in the renormalization
of the chiral loops (while the value of $f$ is by construction
independent on it).  
Only by consistently adopting the same renormalization
prescription the physical quantities remain
invariant. 
By taking $f=0.086\pm 0.013$ GeV we obtain 
the following $\overline{MS}$ results:
\bea
\bar L_4 (M) &=&  (0.4\: ^{+0.4}_{-0.6}) \times 10^{-3} \ ,
\label{barL4M} \\
\bar L_5 (M) &=&  (3.2\pm 0.3) \times 10^{-3} \ ,
\label{barL5M} 
\eea
and 
\bea
\bar L_4 (\Lambda_\chi) &=&  (- 0.7\: ^{+0.4}_{-0.6}) \times 10^{-3} \ ,
\label{barL4Lchi} \\
\bar L_5 (\Lambda_\chi) &=&  (0.1 \pm 0.3) \times 10^{-3} \ .
\label{barL5Lchi} 
\eea
The value of $L_4$ in \eq{barL4M} is consistent with the $\chi$QM 
determination in \eq{L4QM}.
The matching of the expression in \eq{L5QM} with
the result of \eq{barL5M} gives 
a value of $\vev{\bar qq}$ at $\mu=M$  of 
$(-185\pm 10$ MeV)$^3$. 
The larger values of the quark condensate 
that can be extracted at the scale $\Lambda_\chi$
from \eq{barL5Lchi}
are consistent with the range of values of \qq that we obtain
at $\mu=\Lambda_\chi$ from the fit of the $\Delta I =1/2$ rule,
computed in the $\overline{MS}$ scheme.

Even though the actual values for the
parameters \qq and \GG that fit the selection rule
will vary depending on the choice of $f$ in the range
of \eq{f0}, we have verified
that the predictions of the other
observables remain rather stable when varying
the input parameters accordingly.
In particular, the value of $\widehat B_K$
varies by a few percents by changing $f$ 
within the range of \eq{f0}, provided we use the corresponding
values of the input parameters that fit the $\Delta I =1/2$ rule. 
We will therefore take the
central value of $f$ as a fixed input in all numerical estimates.

\section {Hadronic Matrix Elements}

The chiral lagrangian coefficients to order $O(p^2)$
were computed in previous papers: refs.~\cite{I} 
and \cite{IV} for $\Delta S=1$ and $\Delta S=2$ 
transitions respectively.
Here we report the NLO corrections to the matrix elements 
$\langle Q_i \rangle ^{NLO}_{I}$
which are obtained by
means of the building blocks in section 2. 
We define $m_u = m_d\equiv \widehat m$, 
the average of the $u$ and $d$ current quark masses,
consistently to $m_{\pi^\pm} = m_{\pi^0} = m_\pi$.

There is no mass renormalization to this order because all masses
in the matrix elements of the operators $Q_{1-6}$ originate from
the external momenta which are defined to be on the physical mass-shell.
As expected, constant terms 
in the densities, which enter in
the determination of the operators $Q_{5,6}$, cancel in the complete matrix
elements.

The total matrix elements have the form 
\beq
\langle Q_i(\mu) \rangle _{I} \; = \; Z_\pi \sqrt{Z_K}  
\left[\langle Q_i \rangle ^{LO}_{I} + 
\langle Q_i (\mu)\rangle ^{NLO}_{I} \right]  + a_i^I(\mu)\; ,
\label{hme}
\eeq
where $Q_i$ are the operators in \eq{Q1-10},
\beq
\langle Q_i \rangle _{I} \equiv
\langle (\pi \pi)_I | Q_i | K^0 \rangle\ .
\eeq
After including the wavefuntion renormalization, the matrix elements
are expanded to $O(p^4)$, discarding higher order terms.
The functions
$a_i^I(\mu)$ represent the scale dependent meson-loop corrections,
including the mesonic wavefunction renormalization. They are defined
as the isospin projections of the $a_i^{+-}(\mu)$ and $a_i^{00}(\mu)$
corrections computed in ref.~\cite{I}, properly rescaled by factors
of $f_\pi/f$ in order to replace $f_\pi\to f$ in the NLO evaluation.
The scale dependence of the NLO part of the matrix elements 
is a consequence of 
the perturbative scale dependence of the current quark masses
which enter at this order.
 
The LO matrix elements $\langle Q_i \rangle ^{LO}_{I}$
can be found in ref.~\cite{I}, all occurences of $f_\pi$
beeing replaced by $f$.
The NLO $I = 0$ and $2$ contributions to the $K^0\to\pi\pi$ matrix elements  
are given by:
\bea
\langle Q_1 \rangle ^{NLO}_{0} & = & \frac{1}{3} X \left[ 
 \left(-1 + \frac{2}{N_c}\right) \beta 
  - \frac{2}{N_c} \delta_{\vev{GG}} \beta_G \right], \\
\langle Q_1 \rangle ^{NLO}_{2} & = & \frac{\sqrt{2}}{3} X \left[  
 \left(1+\frac{1}{N_c}\right)\beta - \frac{\delta_{\vev{GG}}}{N_c} \beta_G
 \right], \\
\langle Q_2 \rangle ^{NLO}_{0} & = & -\frac{1}{3} X \left[ 
 \left(-2 + \frac{1}{N_c}\right) \beta
 - \frac{\delta_{\vev{GG}}}{N_c}  \left(\beta_G + 3 \gamma_G\right) 
\right], \\
\langle Q_2 \rangle ^{NLO}_{2} & = & \langle Q_1 \rangle ^{NLO}_{2}, \\
\langle Q_3 \rangle ^{NLO}_{0} & = & \frac{1}{N_c} X  \left[ \beta  
 - \delta_{\vev{GG}} \left(\beta_G - \gamma_G\right)
 \right], \\ 
\langle Q_4 \rangle ^{NLO}_{0} & = &\langle Q_2 \rangle ^{NLO}_{0} - 
\langle Q_1 \rangle ^{NLO}_{0} + \langle Q_3 \rangle ^{NLO}_{0}, \\
\langle Q_5 \rangle ^{NLO}_{0} & = &  \frac{2}{N_c} X \; \gamma \\
\langle Q_6 \rangle ^{NLO}_{0} &=&  2 X \left(\gamma
 + \frac{\delta_{\vev{GG}}}{N_c}\gamma_G \right) 
 \eea
where
\beq
X \equiv \sqrt{3} f \left( m_K^2 - m_\pi^2 \right)
\eeq
and the gluon-condensate correction 
$\delta_{\vev{GG}}$ is given by
\beq
\delta_{\vev{GG}} =  \frac{N_c}{2} \frac{\langle
 \alpha_s G G/\pi \rangle}{16 \pi^2 f^4} \, .
\eeq
The quantities $\beta, \beta_G, \gamma$ and $\gamma_G$ are dimensionless 
functions of the mass parameters:
\bea
\beta & = &
 \frac{m_K^2 + 2 m_{\pi}^2}{\Lambda^2_\chi} 
-3 \frac{M}{\Lambda^2_\chi} \left(m_s + 3 \widehat m \right)   
\nnu \\
&& + \frac{m_s - \widehat m}{M} \left(1 - 6 \frac{M^2}{\Lambda^2_\chi}
      \right)  \frac{m_\pi^2}{2\left(m_K^2 - m_\pi^2 \right)}
   + \frac{\widehat m}{M}
   \left( 1 -12 \frac{M^2}{\Lambda^2_\chi}\right) \; , 
\label{beta} \\
\beta_G & = &
 \frac{m_K^2 + 2 m_\pi^2}{6 M^2} 
- \frac{5 m_s + 17\widehat m}{12 M}
- \frac{\left(m_s - \widehat m \right) m_\pi^2}{12 M 
\left(m_K^2 - m_\pi^2\right)} -\frac{\widehat m}{M}  \; , 
\label{betag} \\
\gamma & = & 
\frac{\langle \bar{q} q \rangle}{f^2} \left[
\frac{m_K^2}{ 2 M \Lambda^2_\chi} 
 - \frac{m_K^2 (2 m_s + 6 \widehat m) - m_\pi^2 (m_s + 7 \widehat m)}
{(m_K^2 - m_\pi^2) \Lambda^2_\chi}  \right. \nonumber \\
&& + \left. \frac{2 \widehat m (m_s - \widehat m )}{M (m_K^2 - m_\pi^2)}
\left(2 f_+ - 6 \frac{M^2}{\Lambda_\chi^2} \right) \right] \nnu \\
&& + f_+^2 \left[ - \frac{m_K^2 + m_\pi^2}{ 4 M^2 } 
 +\ \frac{( m_s + 5 \widehat m)}{2 M}  \; + \, 
\frac{2 \widehat m (m_s - \widehat m )}{m_K^2 - m_\pi^2}   \right]
\nonumber \\
&& + \frac{3 M f_+}{\Lambda^2_\chi} \left[ \frac{m_K^2}{2 M} 
 -  \;  \frac{m_K^2 (m_s + 3 \widehat m) - 2 m_\pi^2 (m_s + \widehat m)}
{(m_K^2 - m_\pi^2)}  \right] \; , 
\label{gamma} \\
\gamma_G & = &
\frac{ m_K^2}{m_K^2-m_\pi^2} \frac{m_s-\widehat m}{6 M} \, .
\label{gammag}
\eea
Similarly, for
the $\Delta S=2$ matrix element 
$\langle \bar{K}^0 | Q_{2S} | K^0 \rangle$ we write
\beq
\langle Q_{2S}(\mu) \rangle \; = \; Z_K  
\left[\langle Q_{2S} \rangle ^{LO} + 
\langle Q_{2S}(\mu) \rangle ^{NLO}\right] + a_{2S}(\mu) \; ,
\label{q2s}
\eeq
with
\beq
\langle Q_{2S} \rangle ^{LO} = m_K^2 f^2 \left[1+\frac{1}{N_c}
(1-\delta_{\langle GG \rangle})\right]\ ,
\label{2LO} 
\eeq
and
\bea
\langle Q_{2S} (\mu)\rangle ^{NLO}  =& & m_K^2 f^2 \left\{
\left(1 + \frac{1}{N_c} \right)
\left[ 2 \frac{m_K^2}{\Lambda_\chi^2} + \frac{m_s+\widehat m}{M} \left(
1 -12 \frac{M^2}{\Lambda_\chi^2} \right) \right] \right. \nnu \\   
& & \left.  
- \frac{\delta_{\langle GG \rangle}}{N_c} \left( \frac{m_K^2}{3 M^2} - 
\frac{m_s+\widehat m}{M} \right) \right\} \ ,
\label{2NLO}
\eea
where the scale dependence enters through the perturbative
running of the current quark masses.

The
chiral-loop correction to $\langle \bar{K}^0 | Q_{2S} | K^0 \rangle$, 
including the meson
wave-function renormalization, is given by
\beq
a_{2S}(\mu) =  m_K^2 f_K^2 \left[1+\frac{1}{N_c}
(1-\delta_{\langle GG \rangle})\right]\ r_{2S}(\mu)\ ,
\label{a2s}
\eeq 
where, having normalized the $O(p^4)$ amplitude to $m_K^2 f_K^2$,
\beq
r_{2S}(\mu) = 1.45 + 0.676 \  \ln \mu^2 \, ,
\label{r2s}
\eeq
and $\mu$ is in units of GeV. The given numbers depend only on
the octet meson masses and are obtained for the
values given in appendix C.

By subtracting the $\overline{MS}$ chiral loop renormalization of 
$f^2\to f_K^2$
at the scale $\mu$ one obtains~\cite{IV}
\beq
{\overline r}_{2S}(\mu) = 0.728 + 0.373 \  \ln \mu^2 \ ,
\label{rbar2s}
\eeq
which shows that a large part of the meson-loop correction goes
into the renormalization of $f_K$.

Everywhere, in the NLO as well as in the leading order matrix elements,
the coupling constant $f$ must be understood
as given by the values discussed in section 3.3. 
This replacement in the NLO terms is optional, given that the
difference is of $O(p^6)$. For convenience, in our numerical 
analysis we replace all occurrences of $f$ with its one-loop value.

Another remark concerns the quark condensate which appears in 
the building blocks of the quark densities.
While all factorizable gluonic corrections are absorbed in the physical
definition of the chiral coupling and the quark condensate, there are
mass dependent contributions to the latter
that we have not included in \eq{gamma}.
At the nedeed $O(m_q)$, 
the relation between the condensate $\vev{\bar qq}$  that we
use as a parameter in our analysis, and  the
condensate $\vev{\bar qq}_{m_q}$ calculated at non-zero current
quark masses is given by
\beq
\vev{\bar qq}_{m_q} = 
\vev{\bar qq} + \frac{m_q}{M}\left[\vev{\bar qq} + 2 M f^2 f_+\right]\ . 
\label{qqph}
\eeq
The correction is negligible for 
$q=u,d$ and remains small also for $\vev{\bar ss}$ due to a partial
cancellation between the two terms in square brackets.
In what follows we will always refer to the flavor
independent parameter $\vev{\bar qq}$.

\subsection{Chiral Loop Corrections}

The one-loop chiral corrections $a_\ell(\mu)$
are included at this stage operator by
operator on top of 
the $\chi$QM estimate of the matrix elements, as discussed
in refs. \cite{I,IV}.
Let us recall that throughout our analysis we use dimensional
regularization and the modified minimal subtraction scheme as
it is done for the Wilson 
coefficients. For this
reason the numerical values for the $\Delta S=1$ chiral
loop corrections quoted in ref.~\cite{I} 
differ from those of ref.~\cite{meson} where a non-minimal
subtraction scheme is employed. The two computations are however in 
agreement once the scheme dependence is taken into account.

\section {The $\Delta I =1/2$ Selection Rule}	

As discussed in the introduction,
in order to restrict the possible values of the input parmeters $M$,
\qq and \GG we study the $\Delta I = 1/2$ selection rule
which charactarizes the $I=0$ and $I=2$ amplitudes of
the non-leptonic kaon decays.

Is is convenient to write the generical amplitude 
for a kaon to decay into a two pion
final state of isospin $I=1,2$ as
\beq
{\cal A}_I (K\to\pi\pi) = A_I\ \exp \: i \, (\delta_I)
\label{fullAi}
\eeq
where the phase $\delta_I$ comes from the final
state interactions in the channel $I$.
From \eq{Lquark} and that fact that $\Im\tau \ll 1$ we can write the amplitude
$A_I$ as
\beq 
A_I \simeq \frac{\sum_i C_i(\mu)\ \Re\vev{Q_i(\mu)}_I}{\cos\delta_I}
\label{Ai}
\eeq

Experimentally the phases $\delta_I$ are obtained in terms
of the $\pi$-$\pi$ S-wave scattering lenght~\cite{phase02exp}
at the $m_K$ scale. The values so derived give to a few degrees uncertainty
$\delta_0 \simeq 37^0$ and $\delta_2 \simeq - 7^0$,
thus obtaining with good accuracy
\bea
\cos\delta_0 &\simeq& 0.8 \nnu\\
\cos\delta_2 &\simeq& 1.0\ .
\label{cosdelta02exp}
\eea
As a consequence the $I=0$ amplitude includes a 20\% enhancement
from the rescattering phase.

The existence of $\Im A_I\ne 0$ signal the possible
presence of direct CP violation in the $K\to\pi\pi$ decays, while 
$\omega\equiv |{\cal A}_2|/|{\cal A}_0|\simeq 
\Re A_2/\Re A_0 = 1/22.2$ represents
the $\Delta I= 1/2$ selection rule known experimentally for more than
forty years~\cite{gell-mann pais}.
According to our conventions for the isospin amplitudes we have
$\Re A_0 = 3.3\times 10^{-7}$ GeV and $\Re A_2 = 1.5\times 10^{-8}$ GeV.
Using the notation of \eq{Lquark} we can write the CP conserving
amplitudes as 
\bea
\Re A_0 &=& \frac{G_F}{\sqrt{2}}\frac{V_{ud}\,V_{us}}{\cos\delta_0}
\sum_i z_i\ \Re \vev{Q_i}_0
\nnu \\
\Re A_2 &=& \frac{G_F}{\sqrt{2}}\frac{V_{ud}\,V_{us}}{\cos\delta_2}
\sum_i z_i\ \Re \vev{Q_i}_2\ \left(1 + \Omega_{\eta+\eta'}\right)\ ,
\label{A02}
\eea
where $\Omega_{\eta+\eta'} = 0.25\pm 0.05$ represents the isospin
breaking effect due to the $\pi^0$-$\eta$-$\eta'$ mixing.
For notational convenience henceforth we identify $\Re A_I$ with $A_I$.

For a long time the explanation of the $\Delta I = 1/2$ rule
has been a mystery.
It was found some twenty years ago that perturbative 
QCD qualitatively
worked in the right direction~\cite{wilson,shifman}---even though
 the effect was too small. More recently, it has been shown 
shown that non-factorizable contributions and chiral 
loops also work in the right direction~\cite{BBG,PdeR,meson}. 
In a previous paper
\cite{II}, we argued that the selection rule could be accomodated 
within the context of the $\chi$QM
to $O(p^2)$ with the inclusion of the meson loops. Here we present the  
complete $O(p^4)$ calculation.

To this order in the chiral expansion there are several constraints to be
satisfied. In particular, the stability of the solution
as well as its $\gamma_5$-independence restrict the possible
values of $M$. From Figs. 1 and 2, obtained for the central values
of the parameters given in appendix C, we see that 
in order to keep the numerical results for $A_0$ and $A_2$ within 
20\% from their experimental values
we must constrain $M$ in the range
\beq
M = 180\div 220 \: \mbox{MeV} \label{range-m}
\eeq 
The range thus identified
is in  good agreement with that found by fitting radiative kaon 
decays~\cite{Bijnens}. We notice that
the NLO corrections improve the $\gamma_5$-stability,
especially for $A_2$. Henceforth we will use for the numerical discussion
the HV results. 

\FIGURE{              
\epsfxsize=8cm
\centerline{\epsfbox{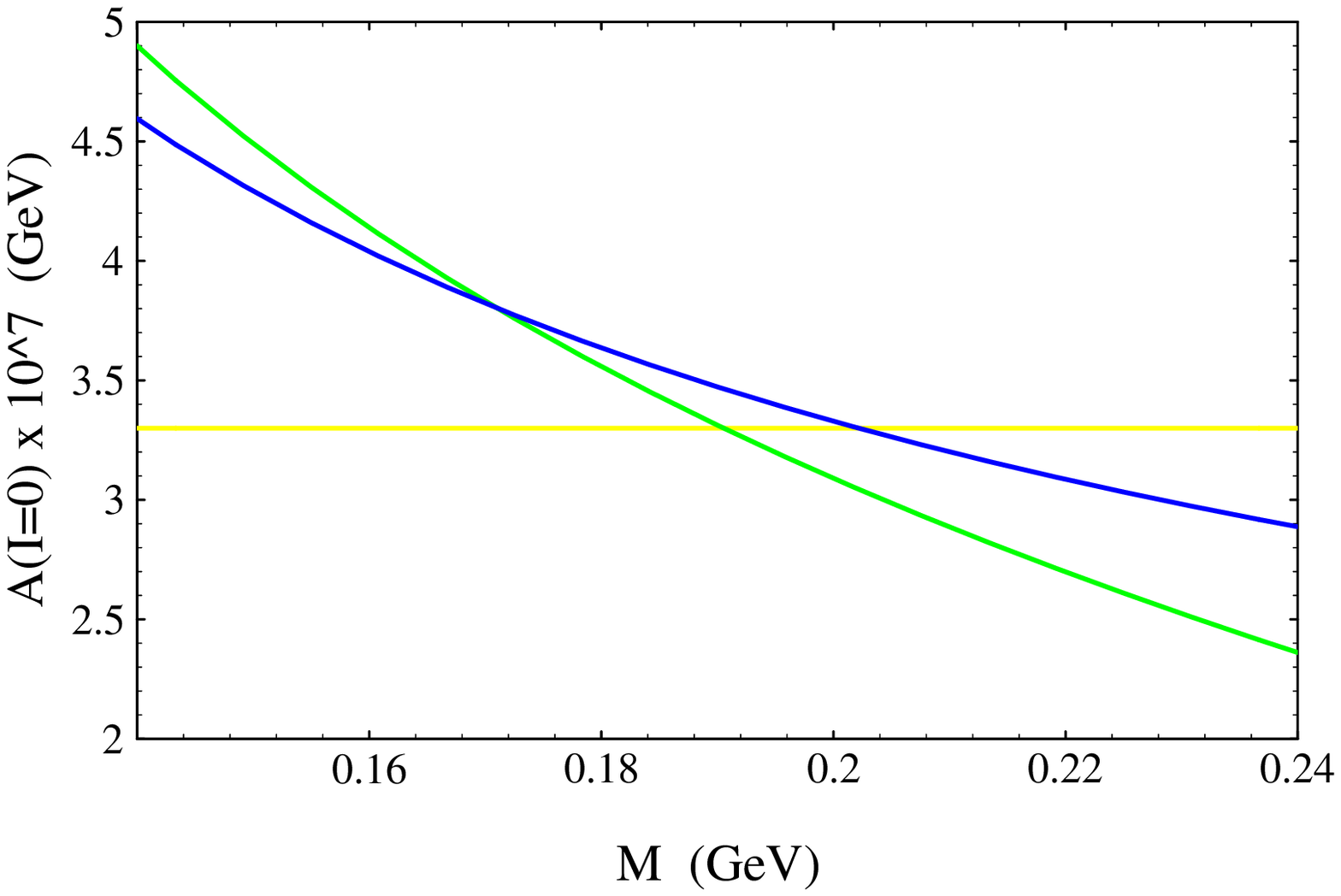}}
\caption{$A_0(K^0\to\pi\pi)$ 
as a function of $M$ in the HV (gray) and NDR (black) schemes respectively.
The horizontal line indicates the experimental value.}}

\FIGURE{              
\epsfxsize=8cm
\centerline{\epsfbox{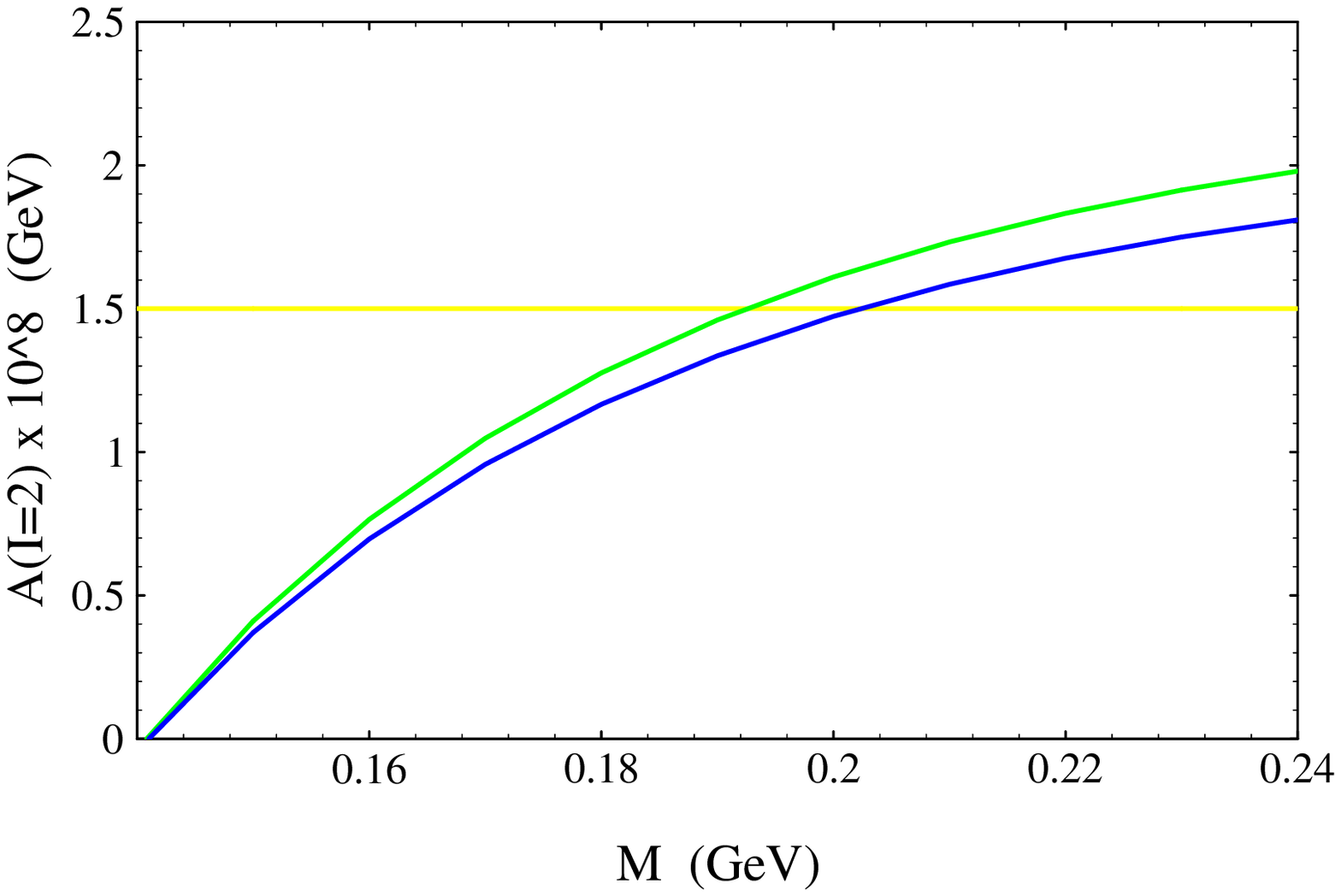}}
\caption{$A_2(K^0\to\pi\pi)$ 
as a function of $M$ in the HV (gray) and NDR (black) schemes respectively.
The horizontal line indicates the experimental value.}}

\FIGURE{              
\epsfxsize=8cm
\centerline{\epsfbox{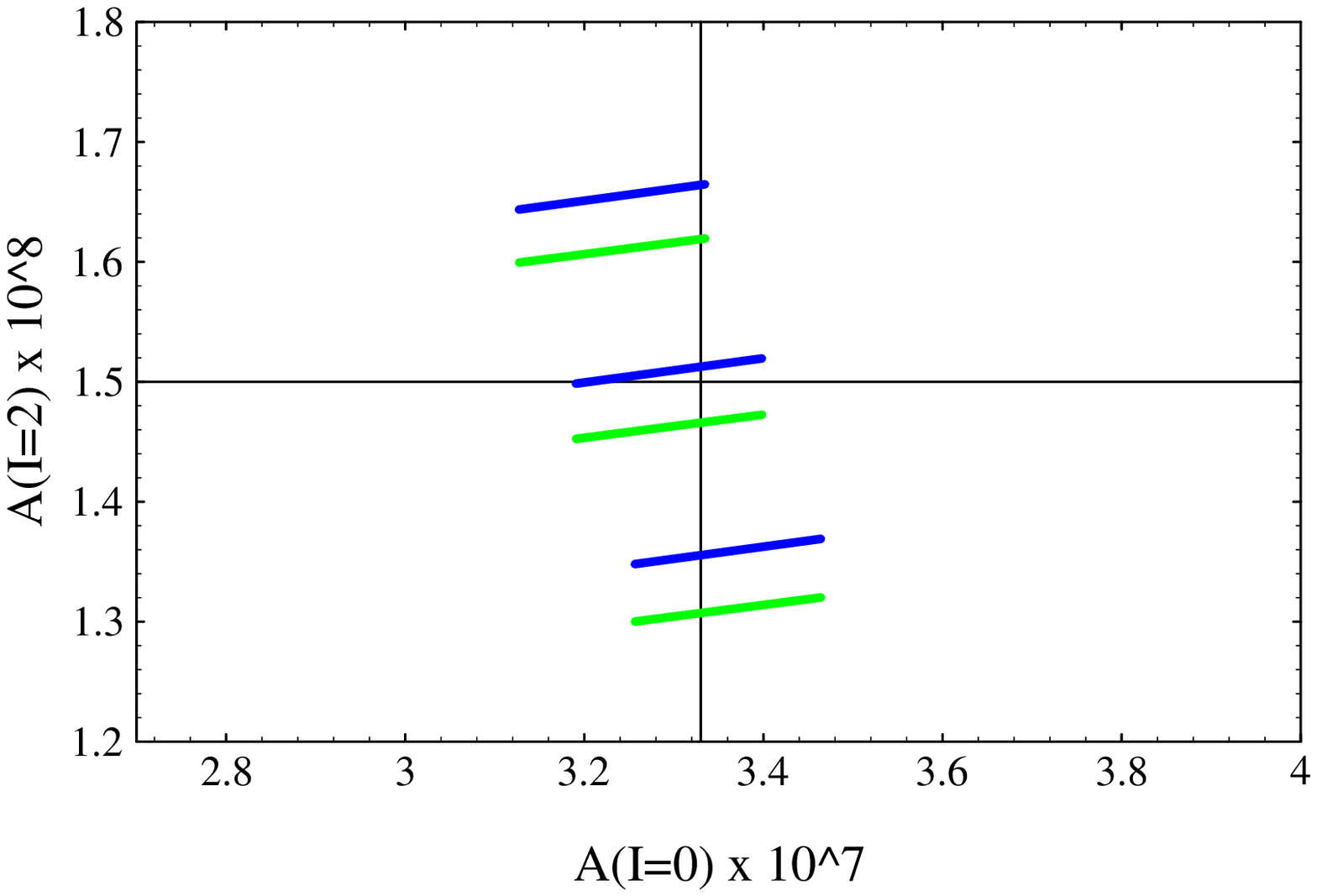}}
\caption{$A_0$ and $A_2$ in units of GeV
as functions of  \GG (vertical spread)
and \qq (horizontal spread) for fixed
 $M=200$ MeV and central value of $\Lambda^{(4)}_{QCD}$.
The cross-hairs indicate the experimental values.
Black and gray lines correspond to $m_s(0.8\ 
\mbox{GeV})=240$ MeV and 
$200$ MeV respectively. The ranges of the gluon and quark condensates
are those discussed in the text. The slight slope in the curves is due to 
the \qq dependence in the electroweak operators.}}

Given the above range for $M$, we
can study the dependence of
the selection rule on the other
$\chi$QM parameters. 
We find that for central values of $M$ and $\Lambda^{(4)}_{QCD}$,
namely $M=200$ MeV and $\Lambda^{(4)}_{QCD}= 340$ MeV,
the $\Delta I=1/2$ rule is reproduced with a $\pm 10\%$ accuracy 
provided
\beq
 \langle \alpha_s GG/ \pi \rangle =  
\left( 334 \pm 4  \:\: \mbox{MeV} \right) ^4\ ,
\label{range-gg}
\eeq
and
\beq
\langle \bar q q \rangle = - \left(240\ ^{+30}_{-10} \:\: 
\mbox{MeV} \right) ^3\ ,
\label{range-qq} 
\eeq
as it is shown in Fig.\ 3.

\FIGURE{              
\epsfxsize=8cm
\centerline{\epsfbox{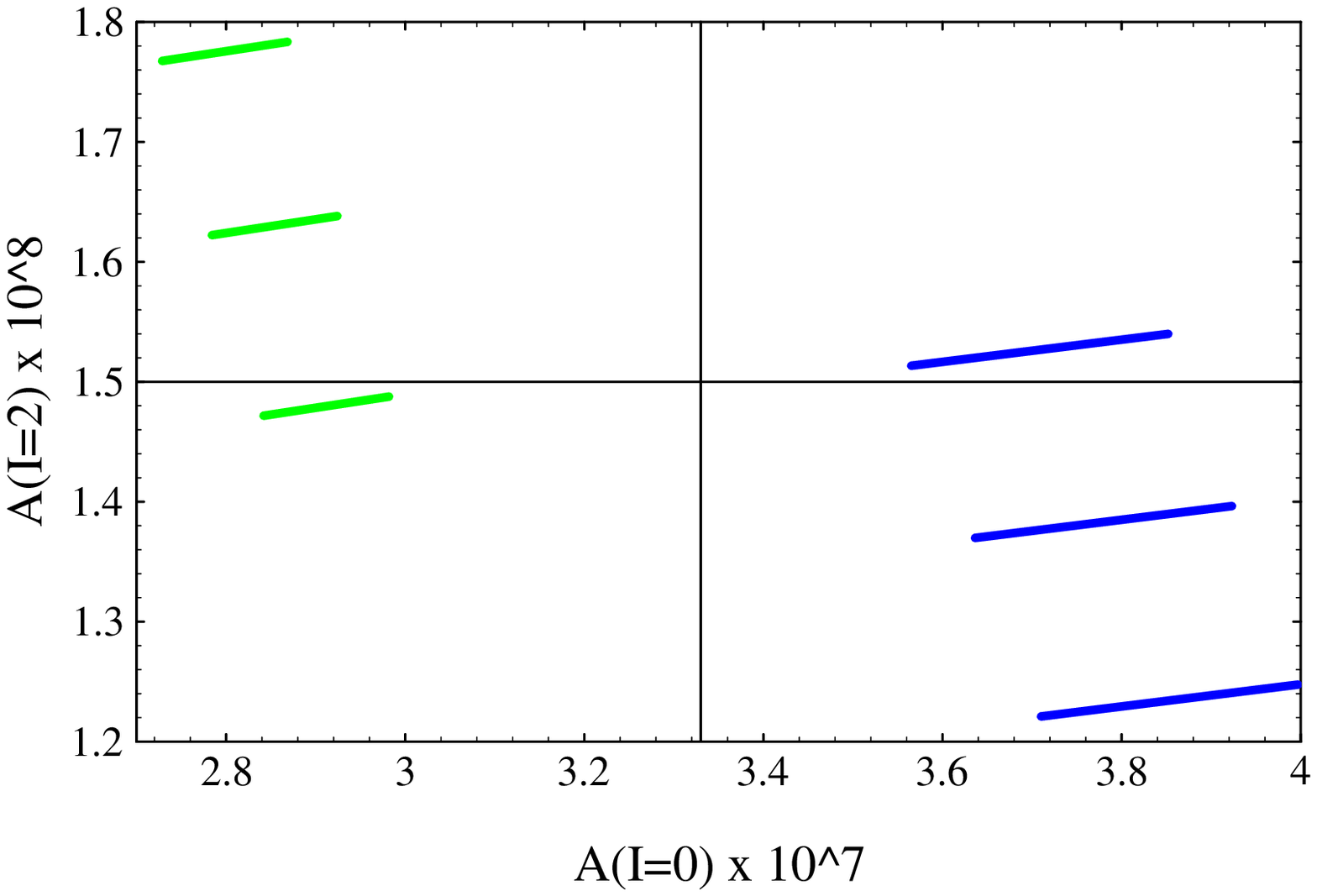}}
\caption{$A_0$ and $A_2$ in units of GeV
as functions of \GG (vertical spread)
and \qq (horizontal spread) in the ranges of \eqs{range-gg}{range-qq}
for $m_s(\Lambda_\chi)=220$ MeV 
and the extreme values of $M$ and $\Lambda^{(4)}_{QCD}$.
Black and gray lines correspond to
$M=205$ MeV ($\Lambda^{(4)}_{QCD}=300$ MeV)
and $197$ MeV ($\Lambda^{(4)}_{QCD}=380$ MeV) respectively.}}

Given the results above, we include the dependence on
$\Lambda_{QCD}$. Requiring that the rule
is finally reproduced with a $\pm 20\%$ approximation, for
$\Lambda^{(4)}_{QCD}$ in the range of \eq{lambdone}
we obtain a sharp constraint on $M$ (Fig. 4):
\beq
M = 200\ ^{+5}_{-3} \:\: \mbox{MeV}\ .
\label{smallrangeM}
\eeq

In order to exhibit the impact of the NLO corrections we have charted
operator by operator the relative weights 
of the leading order computation, of the one-loop chiral
corrections and of the NLO $O(p^4)$ corrections.

\FIGURE{
\epsfxsize=8cm
\centerline{\epsfbox{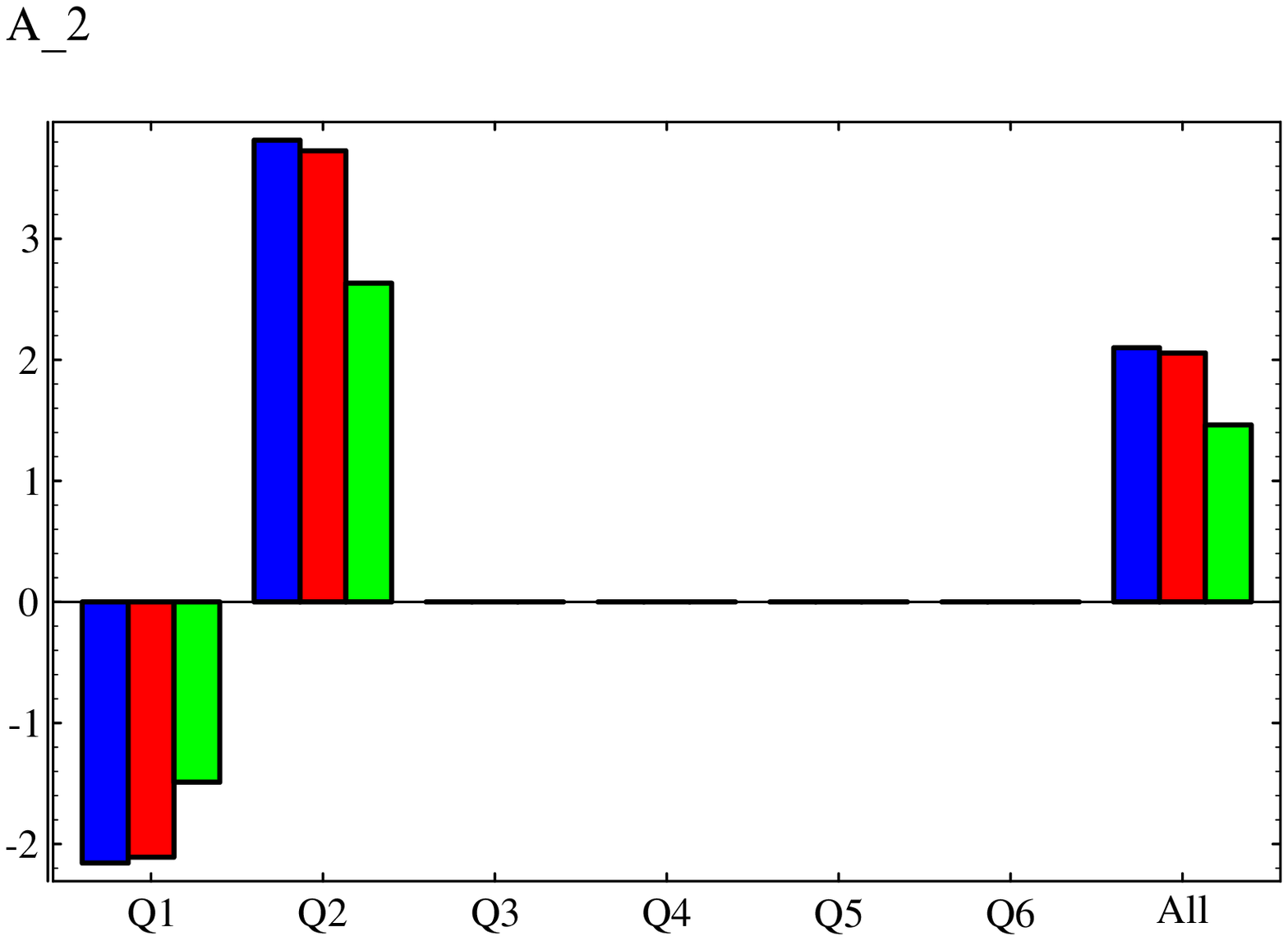}}
\caption{Anatomy of the $A_2$ amplitude in units of $10^{-8}$ GeV
for central values of the input parameters:
LO calculation (black), LO with chiral loops (half-tone),
complete NLO result (gray). 
In the LO case we have taken $f=f_\pi$,
whereas in the remaining hystograms the renormalized value $f=86$ MeV has
been used.}}

\FIGURE{
\epsfxsize=8cm
\centerline{\epsfbox{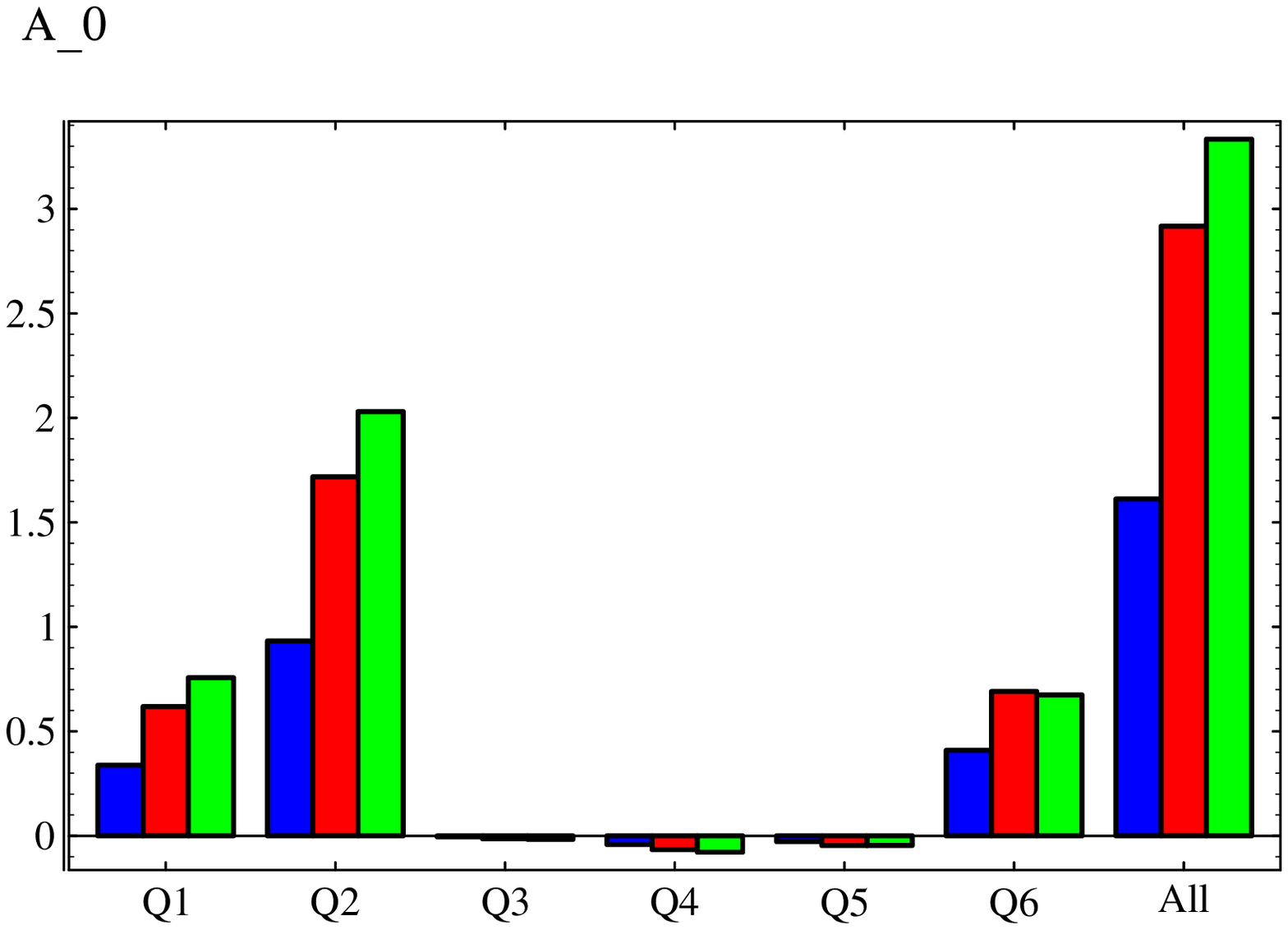}}
\caption{Anatomy of the $A_0$ amplitude in units of $10^{-7}$ GeV
for central values of the input parameters:
LO calculation (black), LO with chiral loops (half-tone),
complete NLO result (gray).
In the LO case we have taken $f=f_\pi$,
whereas in the remaining hystograms the renormalized value $f=86$ MeV has
been used.}}

As one can see from Figs. 5 and 6, the combined
effect of chiral one-loop and NLO $\chi$QM corrections is rather large in the
case of $A_0$ where it leads to an increase of the amplitude by about 50\%,
while for the amplitude $A_2$
chiral loops are negligeable and
$\chi$QM $O(p^4)$ corrections reduce it by 25\%.
It is important to remark that the final $O(p^4)$ result goes in the
direction indicated by the $\Delta I=1/2$ rule, that is, 
of making $A_0$ larger and $A_2$ smaller. 

Another remark regards the gluon penguin operators. Fig.\ 6 makes clear
 that their contribution to the $A_0$ amplitude
 is not as large as often claimed. The $O(p^4)$ correction
is almost completely accounted for by the chiral loops. 

The dependence on $m_c$ is not negligible. For instance, by varying
$m_c(m_c)$ from 1.4 to 1.3 GeV, 
the value of $|\vev{\bar q q}|^{1/3}$ 
required in order to fit the rule increases by about 10\%.

A convenient way of analizing the size of hadronic matrix elements in
different theoretical approaches is via the $B_i$ factors which
quantify the deviation of the hadronic matrix elements
in a particular computation from those obtained in 
the VSA. 
According to the discussion below \eq{fullAi}, we define the $B_i$ as
\beq
B_i^{(0,2)} \equiv \frac{\Re\left[\langle Q_i\rangle_{0,2}^{\rm model}\right]}
{\langle Q_i \rangle _{0,2}^{\rm VSA}} \, ,
\label{Bi}
\eeq
where the VSA matrix elements are by construction real.

In Table 1 we give the $B_i$ coefficients related 
to the operators $Q_{1-6}$ at $\mu =0.8$ GeV.  
The three columns show how the $B_i$ vary from LO 
(which includes the $\chi$QM gluonic corrections) to the complete NLO
result, via the inclusion of meson loops.
Their scale dependence
has been already discussed in ref.~\cite{II}, where also the $\gamma_5$-scheme
dependence was shown. Since the NLO corrections do not affect the latter
we will not repeat the discussion here. 
Anticipating the results of the next section, we have
restricted the values of the input parameters to those required by
the fit of the $\Delta I =1/2$ rule.
\TABLE{
\begin{tabular}{|c||c|c|c|}
\hline
 & {\rm LO} & {+ $\chi$-loops} & {+ \rm NLO} \\ 
\hline
$B^{(0)}_1$  & 4.2 & 7.8 & 9.5\\
\hline
$B^{(0)}_2$  & 1.3 & 2.4  & 2.9\\
\hline
$B^{(2)}_1 =B^{(2)}_2 $  & 0.60 & 0.58 & 0.41\\
\hline
$B_3$ & $-0.62$ & $-1.9$ & $-2.3$\\
\hline
$B_4$ & 1.0& 1.6 & 1.9\\
\hline
$B_5 \simeq B_6$& $1.2\ \div\ 0.72$ & $2.0\ \div\ 1.2 $ & $1.9\ \div\ 1.2$\\
\hline
\end{tabular}
\caption{The $B_i$ factors in the $\chi$QM including meson-loops
and $\chi$QM NLO corrections. We have taken 
the gluon
condensate at the central value of \eq{range-gg}, while the ranges given
for $B_{5-6}$
correspond to varying the quark condensate according to \eq{range-qq}. The 
results shown are given in the HV scheme  for $M = 200$ MeV
and $f=86$ MeV, except for the first column where $f=f_\pi$ has been
used. 
}}

A few comments are in order. The large values for $B^{(0)}_{1,2}$ 
and the small ones
for  $B^{(2)}_{1,2}$ reflect the $\Delta I = 1/2$ selection 
rule, that is well reproduced in our approach.
The decrease in the parameter for the operator $Q_{5-6}$ as we
increase the value of quark condensate is the
consequence of the linear dependence on the quark condensate in the $\chi$QM
with respect to the quadratic one in the VSA. Finally, we confirm 
the rather large and negative value for $B_3$; the effect of such an
unexpected result is discussed in ref. \cite{II}.

We conclude by updating 
in Fig. 7 the ``road'' to the $\Delta I=1/2$ rule
already presented in ref.~\cite{II} were we
plot the values of the amplitudes $A_0$ and $A_2$ as a functions of
the various components
for the central values of the input parameters,
namely $M=200$ MeV, \qq $= (-240 \: \mbox{MeV})^3$, 
\GG $=( 334 \: \mbox{MeV})^4$ and $m_s(0.8$ GeV)$= 220$ MeV. 
This plot includes the NLO $O(p^4)$ corrections (next-to-last step)
that we have here computed. 
Fig. 7 depicts in a suggestive way
the decomposition of the fit into its model-dependent components.

\FIGURE{
\epsfxsize=8cm
\centerline{\epsfbox{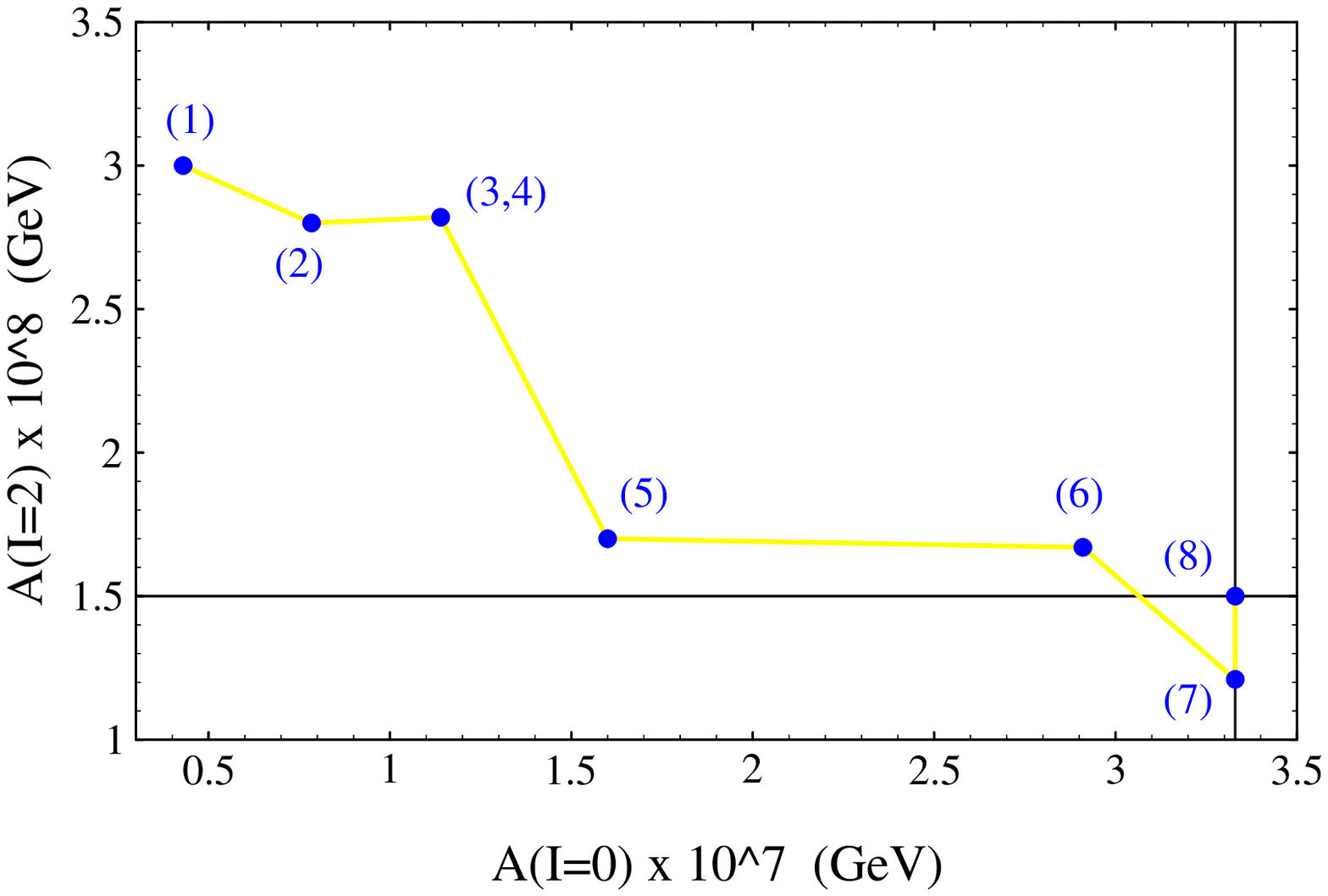}}
\caption{The road to the $\Delta I=1/2$ rule. The numbered points are
discussed in the text.}}

Point (1) represents the result of free quarks (no QCD, only the
operator $Q_2$
is present) combined with the VSA for the matrix element of $Q_2$.
Step (2) includes the effects of perturbative QCD renormalization
on the operators $Q_{1,2}$. 
Steps (3,4) show the effect of including gluon and 
electroweak penguin operators. The latter are responsible 
for the small shift in $A_2$.
Therefore, perturbative QCD brings us from (1) to (4).
A crucial contribution is given by
non-factorizable gluon condensate effects which bring us from (4) to (5)--- 
still remaining at the leading
$O(p^2)$ in the chiral expansion. Moving the analysis at the NLO,
chiral loops computed on the LO chiral
lagrangian lead us from (5) to (6). 
The NLO $O(p^4)$ corrections carachteristic of the $\chi$QM 
calculated in this paper yield the point (7).
Finally, step (8) represents the inclusion of $\pi$-$\eta$-$\eta'$ 
isospin breaking effects which increase $A_2$ by about 25\%. 

Let us remark that the exact match with the experimental values
should not be taken as a theoretical 
prediction but rather as the proof of the reproducibility of the 
experimental rule within our approach.
It should be noted once more that the bulk of the 
effect---up to point (6)---was already included in our previous analysis.

\clearpage
\section {The $K^0$-$\bar{K}^0$ Mixing Parameter $\widehat B_K$ }

The scale-independent
parameter $\widehat B_K$ is defined as the product of
the scale dependent $B_K(\mu)$ parameter in
\beq
\langle \bar{K}^0 | Q_{2S}(\mu) | K^0 \rangle \equiv
\frac{4}{3} f_K^2 m_K^2 B_K (\mu) \, ,
\eeq
and the function $b(\mu)$ introduced in~\eq{b}. 
Its determination is of crucial relevance
in the physics of kaons. We discuss  two independent ways by which
it can be determined within the $\chi$QM approach.

\subsection{$\widehat B_K$ from a Direct Computation} 

A ``model independent'' estimate of $\widehat B_K$ can be
obtained by considering the relationship between
the $\Delta S = 2$ matrix element and
that of the $\Delta S =1$ and $\Delta I = 3/2$ amplitude
${\cal A}(K^+\to\pi^+\pi^0)$ on the basis of the chiral symmetry~\cite{DGH}
\beq
\frac{4}{3} f_K^2 m_K^2 \widehat B_K =
\frac{\sqrt{2}}{G_F}\frac{f_\pi}{V_{us}^*V_{ud}}\frac{m_K^2}{m_K^2-m_\pi^2}
\frac{b(\mu)}{z_1(\mu)+z_2(\mu)}\ {\cal A}(K^+\to\pi^+\pi^0)\ .
\label{bkdi}
\eeq

Having a model that reproduces the experimental $I=2$ amplitude, 
we must subtract from the expression of
${\cal A}(K^+\to\pi^+\pi^0)$ all the chiral symmetry breaking
components due to chiral loops, NLO corrections, 
electroweak penguins and $\pi-\eta$ mixing.
In this way one obtains in the $\chi$QM approach, on the basis of chiral 
symmetry alone, the following prediction~\cite{II}: 
\beq
\widehat B_K =\frac{3}{4} b\left(\mu\right) \frac{f_\pi^2}{f_K^2} 
\left[1+\frac{1}{N_c} \left(1-\delta_{\langle GG \rangle} \right)\right] \ , 
\label{BK0} 
\eeq
which includes the non-factorizable gluonic corrections.

If we choose for the gluon condensate
the value \GG = (334 MeV)$^4$, 
which gives the best fit of ${\cal A}(K^+\to\pi^+\pi^0)$ at $O(p^4)$, 
we obtain at $\mu = 0.8$ GeV and in the HV scheme the prediction 
\beq
\widehat B_K \simeq 0.47\ , 
\label{BKp2}
\eeq
to which all the specific chiral symmetry breaking corrections must
be added. Notice that
this value is about two times smaller than the LO result
depicted in Fig. 8. This is 
due to the presence of $f_\pi$ in \eq{BK0}
as derived from \eq{bkdi}, both in the factor $f_\pi^2/f_K^2$ and in
$\delta_{\langle GG \rangle}$. 
In our leading order estimate
we have instead taken $f=f_K$ ($f_\pi^2/f_K^2\to 1$), 
which we believe is the consistent choice
for the tree level evaluation of the $K^0$-$\bar K^0$ matrix element
(as a matter of fact it minimizes the NLO corrections).
Thus the leading order $\chi$QM component of the $O(p^4)$ fit amounts to
\beq
\widehat B_K \simeq 0.9\ . 
\label{BKp2QM}
\eeq

\FIGURE{
\epsfxsize=8cm
\centerline{\epsfbox{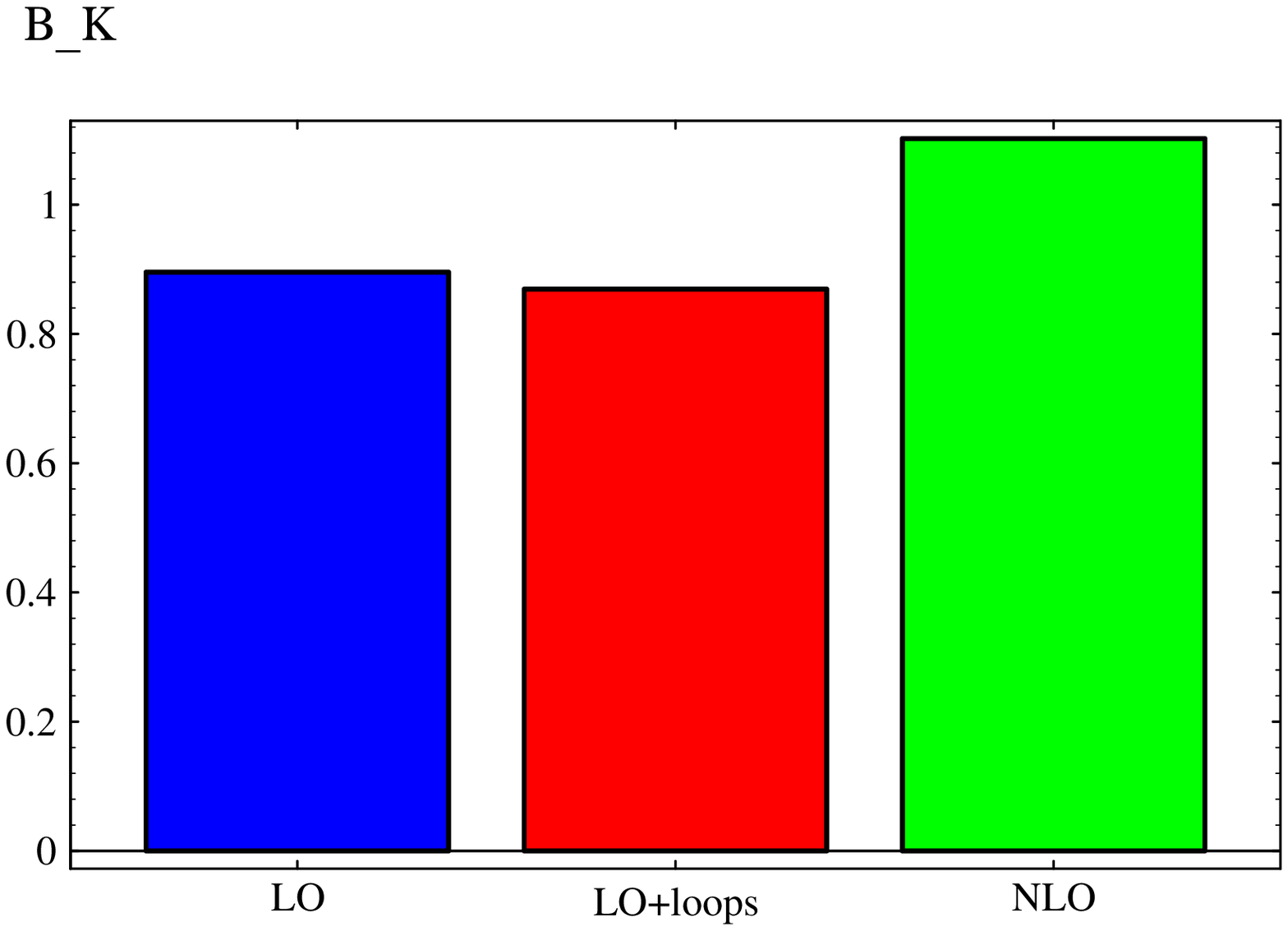}}
\caption{Anatomy of $\widehat B_K$ for central values
of the input parameters in the HV scheme: 
leading order result (LO), 
with chiral loops (LO + loops), and complete NLO result (NLO).
In the LO case we have taken $f=f_K$,
whereas in the remaining hystograms the renormalized value $f=86$ MeV has
been used.}}

Adding to \eq{BK0} the meson-loop corrections
($f_\pi\to f$) and 
the NLO corrections of \eq{2NLO}, we find the following formula
\bea
\widehat B_K = \frac{3}{4} b\left(\mu\right) 
\frac{f^2}{f_K^2}
&&\left[ 
  \left(1 + \frac{1}{N_c}\right)\ 
\left(1 + \rho(\mu) + \frac{f_K^2}{f^2}\ r_{2S}(\mu) \right)\right. \nnu \\
&&\left.  - \frac{\delta_{\langle GG \rangle}}{N_c} \  
\left(1 + \lambda - \rho(\mu) + \frac{f_K^2}{f^2}\ r_{2S}(\mu) \right) 
\right] \ ,
\label{BKNLO} 
\eea
where the chiral loop effects are included
through the function $r_{2S}(\mu)$ of \eq{r2s}.
The functions $\rho$ and $\lambda$ are given by
\bea
\rho(\mu) &=&  \frac{m_s(\mu) + \widehat m(\mu)}{M}
              \left( 1 - 6 \frac{M^2} {\Lambda^2_\chi}\right)\ ,
\label{rhomu} \\
\lambda &=&  \frac{m_K^2}{3 M^2} 
              \left( 1 - 6 \frac{M^2} {\Lambda^2_\chi}\right)\ .
\label{lambda}
\eea
A simpler expression of \eq{BKNLO} is obtained by replacing
$f$ by $f_K$ via the proper $\chi$QM and meson loop renormalizations
discussed in subsections 3.2 and 3.3.
We then obtain
\beq
\widehat B_K = \frac{3}{4} b\left(\mu\right) 
\left[ 
  \left(1 + \frac{1}{N_c}\right)\ 
\left(1 + {\overline r}_{2S}(\mu) \right)
 - \frac{\delta_{\langle GG \rangle}}{N_c} \  
\left(1 + \lambda - 2\ \rho(\mu) + {\overline r}_{2S}(\mu) \right) 
\right] \ ,
\label{BKNLOsimple} 
\eeq
where ${\overline r}_{2S}(\mu)$ is given in \eq{rbar2s}.
This corresponds to eq. (3.17) in ref.~\cite{IV} with the
addition of the $O(p^4)$ $\chi$QM corrections here computed.

\FIGURE{
\epsfxsize=8cm
\centerline{\epsfbox{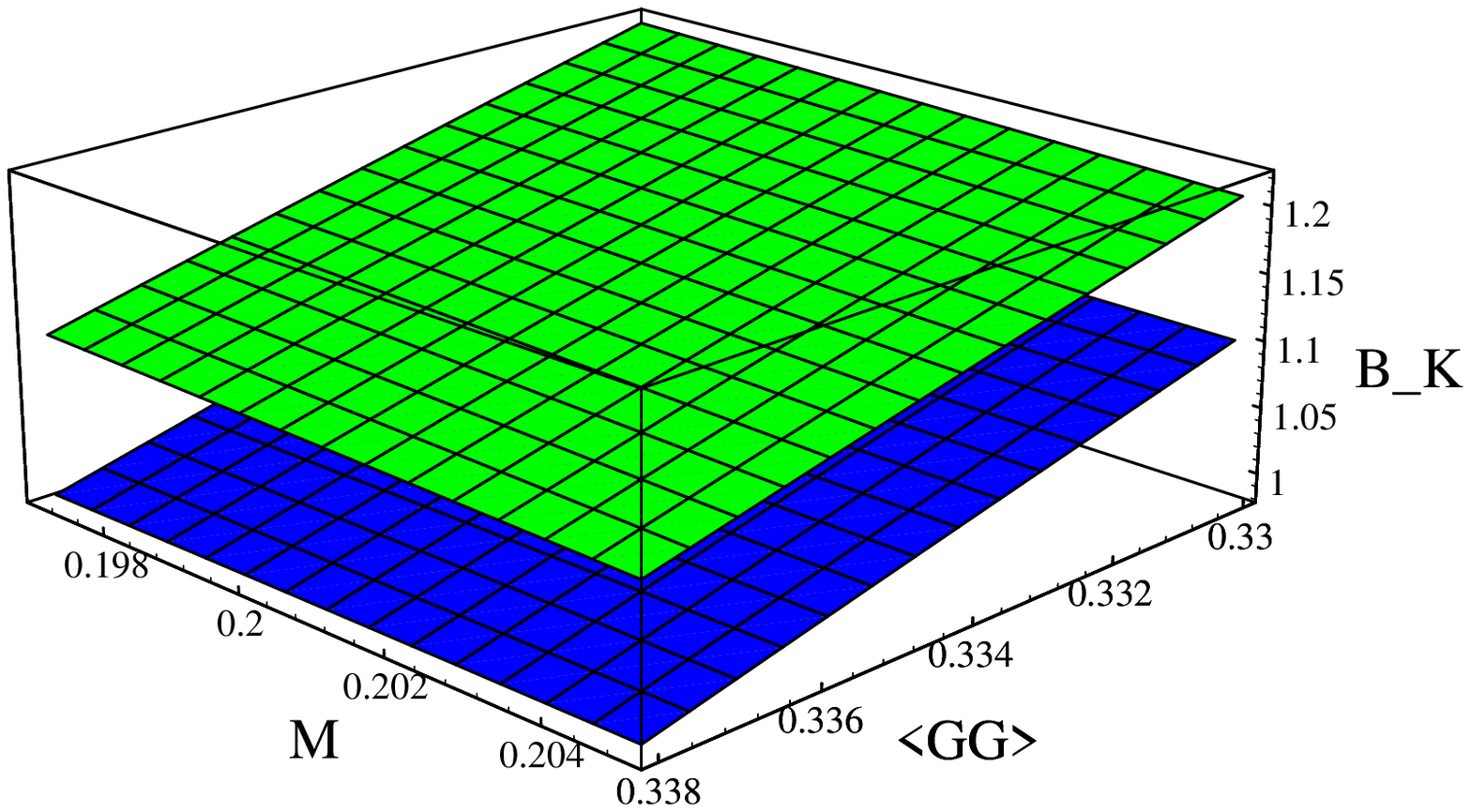}}
\caption{The scale independent parameter $\widehat B_K$ as a function
of  $M$ and ${\langle GG \rangle}\equiv$\GG\hspace{-0.3cm}~$^{1/4}$ (in GeV)
for $m_s(0.8\ \mbox{GeV}) = 200$ (lower surface) and $240$ MeV
(upper surface).}}

By varying \GG in the range of \eq{range-gg}, $M$ in the
range of \eq{range-m} and $\Lambda^{(4)}_{\rm QCD}$ in that given in
\eq{lambdone}, we find
\beq
\widehat B_K = 1.1 \pm 0.2 \, ,  \label{bk}
\eeq
where the error includes a 8\% uncertainty due to the 
$\gamma_5$-scheme dependence
of $b(\mu)$ at $\mu=0.8$ GeV. 

In Fig.\ 8 we have charted 
the relative weight of the LO, chiral loop corrections
and NLO contributions for central values of the input parameters..
The renormalization of $\widehat B_K$ from 0.9 to 1.1 amounts to about 20\%
and it is due to the effect of the 
NLO corrections that increase the value
of the parameter (having chosen $f=f_K$ in the LO estimate
minimizes the $O(p^4)$ corrections). 
This pattern differs from that found in
ref. \cite{Bruno} where the NLO corrections
act in the opposite direction, reducing the final value. 
However one has to keep in mind that the anatomy of the various contributions
depends on the subtraction prescription used in the renormalization
of the meson loops.

To give a direct impression of the variation of  $\widehat B_K$ as we
vary the input values, we show in Fig.\ 9 the overall dependence on 
\GG, $M$ and $m_s$.
We recall that the NLO result for $\widehat B_K$ is very little
dependent on the value of $f$ used, in the range of \eq{f0}, provided
a corresponding fit of the $\Delta I =1/2$ rule is performed.

This computation has a systematic scale uncertainty---already
present at the leading order, and
discussed in ref.~\cite{IV}---due to the poor matching between
the long- and the short-distance computations,
at variance with the cases of the
amplitude $A_0$ and $\varepsilon'/\varepsilon$~\cite{III}, 
where the scale dependence of the Wilson coefficients
is drastically reduced by the meson loops.
This is a consequence of the chiral-loop scale dependence of
$a_{S2}(\mu)$ that adds coherently to
the running of $b(\mu)$.
In this respect higher order corrections do help,
since the running of $m_s$ improves the scale stability.
Changing the matching scale from 0.8 to 1 GeV 
has the effect of increasing the value of $B_K$
by about 15\%.

As a final remark, 
in the chiral limit where $m_q =0$ (but $m_K \neq 0$), 
the NLO corrections change sign
and actually decrease the final value of the parameter; in this case, we find
\beq
\widehat B_K (m_q=0) = 0.47 \pm 0.11 \, .
\eeq
This dramatic change qualitatively agrees
with what found in ref. \cite{Bijnens-Prades}. 

\subsection{$\widehat B_K$ from the $K_L$-$K_S$ Mass Difference}

An alternative and independent way of deriving $\widehat B_K$ 
is by computing the $\bar K_L$-$K_S$
 mass difference
\beq
\Delta M_{LS} \equiv m_L - m_S \, ,
\eeq
and then comparing with the experimental value
\beq
\Delta M_{LS}^{\rm exp} = (3.510 \pm 0.018) \times 10^{-15} \; \mbox{GeV} \, .
\label{DMLSexp}
\eeq 
This 
computation is made of two parts:
\beq
\Delta M_{LS} = \Delta M_{SD} + \Delta M_{LD} \, ,
\label{DMLS}
\eeq
where $\Delta M_{SD}$ accounts for the short-distance contribution due to the
$\Delta S=2$ effective lagrangian in \eq{lags2} and
$\Delta M_{LD}$ for
the long-distance effects generated by the double insertion of the $\Delta S=1$
chiral lagrangian~\cite{IV}.

By combining \eq{DMLSexp} and \eq{DMLS}
we can extract the value of
$\widehat B_K$ as
\beq
\widehat B_K = b(\mu)\frac{\Delta M_{LS}^{\rm exp} -  
\Delta M_{LD}}{C_{S2}(\mu)\ (4/3) f_K^2 m_K^2 } \, ,
\label{BK_DM}
\eeq
where $C_{S2}(\mu)$ is defined in \eq{lags2}.

In a previous paper~\cite{IV} we have calculated 
within the $\chi$QM the long-distance contributions
to $\Delta M_{LS}$ and shown that the net result has the sign
opposite to that of the dominant short-distance component,
so that the long-dis\-tan\-ce corrections to
the mass difference actually ma\-ke the prediction 
smaller\footnote{In the present estimate of $\Delta M_{LD}$ we
have not used the Gell-Mann-Okubo (GMO) relation 
(as we did in ref.~\cite{IV}). 
As a consequence $\Delta M_{LD}$ includes also the
single-pole diagrams that vanish in the limit in which 
the GMO relation holds.}.

According to \eq{BK_DM}, the value of
$\widehat B_K$ that fits the experimental value of $\Delta M_{LS}$ is
found to be
\beq
\widehat B_K = 1.2 \pm 0.1\ , 
\label{bkm}
\eeq
where the error is obtained
by a flat span of the whole range of the input values 
($\Lambda^{(4)}_{\rm QCD}$ included). The 
effect of changing the matching scale from 0.8 GeV
to 1 GeV reduces the absolute value of $\Delta M_{LD}$, corresponding to 
a decrease of $B_K$ by less than 10\%, which is within the error given. 

The overlap of the two results (\ref{bk}) and (\ref{bkm})
is rather encouraging and should dispell
the concern about the rather
large value of $\widehat B_K$ we find.

\acknowledgments{We thank V. Antonelli for his participation to
the initial part of this work. Work 
partially supported by the Human Capital and Mobility EC program under 
contract no. ERBCHBGCT 94-0634. JOE thanks SISSA for its hospitality. }

\appendix

\section{$O(m_q^2)$ Contributions to $\vev{Q_6}$}

As pointed out in ref. \cite{bef0} the matrix element of $Q_6$ can be written
as
\bea
\langle \pi^+ \pi^-|Q_6| K^0 \rangle \, &=&
  2 \, \langle \pi^-|\overline{u}\gamma_5 d|0 \rangle
\langle \pi^+|\overline{s} u |K^0 \rangle
- 2\ \langle \pi^+ \pi^-|\overline{d} d|0 \rangle  \langle 0|\overline{s} 
\gamma_5 d |K^0 \rangle
\nonumber \\
& & +\ 
2\ \left[ \langle 0|\overline{s} s|0 \rangle \, - \, \langle 0|\overline{d}
d|0 \rangle
\right] \,  \langle \pi^+ \pi^-|\overline{s}\gamma_5 d |K^0 \rangle \, .
\label{MQ6}
\eea
 The last line in \eq{MQ6}
was previously neglected because 
$\langle \pi^+ \pi^-|\overline{s}\gamma_5 d |K^0 \rangle$
is zero to second order in  momentum. However, there are
$O (m_q^2)$ terms, obtained when 
$\left[ \langle 0|\overline{s} s|0 \rangle\right. \ -$
$\ \left.\langle 0|\overline{d} d|0 \rangle \right] \, \sim \, (m_s - m_d)$ 
is combined with $O(m_q)$ contributions from
$ \langle \pi^+ \pi^-|\overline{s}\gamma_5 d |K^0 \rangle $, which 
a priori have to be included in a NLO analysis.
A direct calculation gives 
\bea
 \langle \pi^+ \pi^-|\overline{s}\gamma_5 d |K^0 \rangle \, = \,
-i\sqrt{2} \left\{ \frac{ \langle \overline{q} q \rangle}{3f^2} \, + \, 
\frac{f_+}{f}\left[\frac{1}{3}(m_s + m_d) + (m_d + m_u) \right] \right\}
\; ,
\label{Kpipi}
\eea
and we must worry about terms proportional to $m_s$ that can be numerically
relevant.
However, some parts of this contribution can be rotated away, as we shall explain below. In fact, only the last term proportional to $m_d + m_u$ corresponds
to a physical contribution.

The quark condensate term in \eq{Kpipi} combined with
the $m_s-m_d$ component of 
$ \langle 0|\overline{s} s|0 \rangle \, -
 \, \langle 0|\overline{d} d|0 \rangle $ 
correspond to a chiral lagrangian term proportional to
\begin{equation}
 \Tr \Bigl[ \Sigma^{\dagger} \, {\cal{M}}_{q} \, \lambda_- +  \lambda_- \,
\label{14}
{\cal{M}}_{q}^{\dagger} \,  \Sigma \Bigr] \; .
\label{masslag}
 \end{equation}
where $\lambda_\pm = (\lambda_6 \pm \lambda_7)/2$ are the combination of 
Gell-Mann matrices projecting out $\Delta S = \pm 1$ transitions.
This term  contains $K^0$ to vacuum transitions and can therefore 
be rotated away~\cite{bef0} in agreement with the FKW theorem~\cite{FKW}. 
In general, 
\bea
\langle 0 |Q_6| K^0 \rangle \, &=&
2 \, \left[ \langle 0|\overline{s} s|0 \rangle \, - \, \langle 0|\overline{d}
d|0 \rangle \right] \,  \langle 0 |\overline{s}\gamma_5 d |K^0 \rangle \, ,
\label{MQ6V}
\eea
and we have to find out whether there are also terms $O(m_q^2)$
which can be rotated away. To investigate this we do a calculation in the
rotated picture of the $\chi QM$. Within this picture~\cite{bef0},
\bea
Q_6 \, = \, - \, 8 \, (F_{(-)})_{\alpha \beta}
 \, (\overline{\cal{Q}}_L)_{\alpha} ({\cal{Q}}_R)_{\delta}
(\overline{\cal{Q}}_R)_{\delta} ({\cal{Q}}_L)_{\beta} \, ,
\eea
where $F_{(-)} \; = \,  \xi \, \lambda_- \xi^{\dagger}$,
and the Greek letters are flavor indices for the constituent quark fields
${\cal{Q}}_L = \xi q_L$ and ${\cal{Q}}_R = \xi^\dagger q_R$. Here 
$\xi \xi = \Sigma$, $L = (1 - \gamma_5)/2$, and $R = (1 + \gamma_5)/2$.
 To obtain the
 contribution $O(m_q^2)$ from $Q_6$ to the chiral lagrangian we have
 to contract the 
quark fields in $Q_6$ to the product of two quark loops with two mass 
insertions. Within the rotated picture, the mass term reads
${\cal{L}}_{mass} \, = \, - \, \overline{{\cal{Q}}}
 \widetilde M_q  {\cal{Q}}$,
where
\bea
 \widetilde M_q   \; \equiv  \; \xi^{\dagger} \, {\cal{M}}_{q}
  \xi^{\dagger} \, L  \;   +  \;   \xi \,
{\cal{M}}_{q}^{\dagger} \,  \xi \, R \, \equiv \,
 \widetilde M_q^S  + \widetilde M_q^P \, \gamma_5 \; . 
\label{mass}
\eea
where   $\widetilde M_q^S$ and  $\widetilde M_q^P$ are defined in an
 obvious manner  to be independent of $\gamma_5$, and  ${\cal{M}}_{q}$
is the current mass matrix. There are three diagrams.
First, there is a diagram (A) with two mass insertions in the loop 
involving the
right-handed vertex, and no mass insertion in the second quark loop (which
corresponds to the quark condensate divided by two), giving the contribution :
\begin{equation}
 {\cal{L}}_A \, = \, \frac{1}{2} \langle \bar{q} q \rangle \, (-i N_c) \,
\Tr  \Bigl[ F_{(-)} \, R \, S(p) \, \widetilde M_q \, S(p) \, \widetilde M_q
\, S(p) \Bigr]  \,  ,
\label{LA}
\end{equation}
where $S(p) = ( \gamma \cdot p - M)^{-1}$, and the trace is both in 
flavor and Dirac spaces. Further there is a diagram (B) with both mass
 insertions in the loop with the left-handed vertex, and no mass insertion
 at the right-handed vertex, and finally a diagram (C) with one mass 
insertion in both loops. The sum of the chiral lagrangians for the diagrams
A,B,C are a linear combination of the terms
\begin{equation}
\widehat  {\cal{L}}_{XY} \, =  \, 4 \,
\Tr  \Bigl[ F_{(-)} \, \widetilde M_q^X \,  \widetilde M_q^Y  \Bigr]  \,  ,
\label{LXY}
\end{equation}
with $(XY) = (SS), (PP), (SP), (PS)$. We have found that the term 
\begin{equation}
\widehat {\cal{L}}_{PP}  \; = \;
 \Tr \Bigl[\lambda_-  {\cal{M}}_{q}^{\dagger}  \Sigma \,
{\cal{M}}_{q}^\dagger  \Sigma   \,  +
  \, \lambda_-  \, \Sigma^\dagger \, {\cal{M}}_{q} \Sigma^\dagger 
 {\cal{M}}_{q} \, - \, \lambda_-  \, \Sigma^\dagger  {\cal{M}}_{q}
 \, {\cal{M}}_{q}^\dagger  \, \Sigma  \Bigr]  \; ,
\label{LPP}
\end{equation}
contains no $K$ to vacuum  transitions and will correspond to physical effects.
In the total lagrangian $O(m_q^2)$ we also have terms involving
the combinations
\begin{equation}
\widehat {\cal{L}}_{SS}  \; = \; \widehat {\cal{L}}_{PP} \; + \; 2 \, 
\delta \widehat {\cal{L}} \; , \; \;
\widehat {\cal{L}}_{SP} \, - \, \widehat {\cal{L}}_{PS} \; = \; 2 \,
\delta \widehat {\cal{L}} \; \; ,
\label{Lhats}
\end{equation}
where
\begin{equation}
\delta \widehat {\cal{L}}  \; = \;
 \Tr \Bigl[ \lambda_-  \, \Sigma^\dagger  {\cal{M}}_{q}
 \, {\cal{M}}_{q}^\dagger \, \Sigma \Bigr]  \; ,
\label{dL}
\end{equation}
This term has a $K^0$ to vacuum transition, and can be rotated away.
Therefore, we obtain the total contribution $O(m_q^2)$ from $Q_6$
as
\begin{equation}
 {\cal{L}}_A \, + \, {\cal{L}}_B \, + \,  {\cal{L}}_C  \; = \;
\left[\frac{\langle \bar{q} q \rangle }{M} \left(2 f_+ \, - \, 
6 \frac{M^2}{\Lambda_\chi^2}\right) \, + \, f_+^2 f^2 \right] 
f^2 \widehat {\cal{L}}_{PP}
 \; + \; G_\delta  \, 
\delta \widehat {\cal{L}}  \; \; ,
\label{TotL}
\end{equation}
where the term proportional to $\delta \widehat {\cal{L}}$ has no 
physical consequences. 

What we found is that the terms in \eq{Kpipi}
containing the numerical factor 1/3 correspond to $\delta \widehat {\cal{L}}$.
Thus, the numerically most important terms $\sim m_s^2$ are rotated away.
Adding the last  term proportional to $m_u+m_d$ in
(\ref{Kpipi}) to what we obtain from the building blocks in section 2,
we obtain the result contained in the quantity $\gamma$ in (\ref{gamma}), which
agrees with the result in \eq{TotL}.

\section{Computing $L_5$ and $L_8$ in the Strong Sector}

The higher order (renormalized) counterterms in the strong
chiral lagrangian can be determined by using the $\chi$QM.
This has been done in 
 \cite{QM1}  using the technique of path
integral; here we  re-derive some of them in a different
and simpler way.

The method consists in
 computing the Green functions with mesons  as external states by
using the $\chi$QM lagrangian 
given in \eq{M-lag} and then to compare it with the ones computed
using the strong chiral 
lagrangian\footnote{It should be understood that the computation
includes only tree diagrams 
induced by $O (p^2$) and  $O(p^4$) terms in the strong chiral
lagrangian, since the chiral loop contributions are common
for the two approaches.}. From this comparison 
the higher order counter terms appearing in the strong chiral
lagrangian can be determined. 
This is also the way  we have followed  in our previous work
\cite{I} to determine the $\Delta S =
1$ weak chiral lagrangian up to $O (p^2)$.

As an example, we present the determination of $L_5$ and $L_8$. 
The same method
can be extended to determine the other higher counter
terms in the strong chiral lagrangian. 

The propagator of the meson field 
can be represented in momentum space as,
\beq
\frac{i}{p^2 - m_0^2 - \Sigma\left(p^2\right)},
\label{prop}
\eeq
where $\Sigma\left(p^2\right)$ is 
the self-energy of the meson and $m_0$ is its tree-level (bare) mass.
 The self energy $\Sigma\left(p^2\right)$ may be 
organized as an expansion around 
an arbitrary point in the momentum space  $\mu$,
\beq
\Sigma\left(p^2\right) = \Sigma\left(\mu^2\right) + \left(p^2 - \mu^2\right) 
\Sigma' \left(\mu^2\right) + \tilde{\Sigma}\left(p^2\right),
\label{self}
\eeq
where the prime indicates the derivative with respect to $p^2$ and
$\tilde{\Sigma}\left(p^2\right)$ denotes the other terms in the expansion.
Fixing $\mu$ to be the on-shell momentum, 
and defining the physical mass as the pole of the
propagator, \eq{prop} and \eq{self} lead to 
\beq
m_0^2 + \Sigma \left(m^2\right) = m^2 ,
\label{phmass}
\eeq
which relates the physical mass and the bare mass.
The wave-function renormalization  
is then given by
\beq
Z = \left[1 - \Sigma'\left(m^2\right)\right]^{-1} = 
1 + \Sigma'\left(m^2\right) + \left[\Sigma'\left(m^2\right)\right]^2
+ \cdots \, .
\label{zren}
\eeq

Computing the two-point function  by means of the $\chi$QM, we get
\bea
\Sigma \left(p^2\right) 
&=& 
- \frac{p^4}{\Lambda_\chi^2} 
+ 6 \frac{M \left(m_u + m_d \right)}{\Lambda_\chi^2}p^2 
- (m_u+m_d)^2 \nnu \\
&& 
- \frac{p^6}{10 M^2 \Lambda_\chi^2}
+ \frac{\left( m_u + m_d \right)}{M \, \Lambda_\chi^2} p^4 
- 6 \frac{m_u  m_d}{\Lambda_\chi^2} p^2 
+ O\left(m_q^3\right)
\ .
\eea
 From \eq{phmass} and \eq{zren}, we obtain
\bea
Z_\pi\ =\ 1 
&&\hspace{-0.5cm} 
-\ 2 \frac{m_\pi^2}{\Lambda_\chi^2} 
+ 6  \frac{M \left( m_u + m_d \right)}{\Lambda_\chi^2}
- 6 \frac{m_u m_d}{\Lambda_\chi^2}
+ 2  \frac{\left( m_u + m_d \right) m_\pi^2}{M\ \Lambda_\chi^2}
- \frac{3 m_\pi^4}{10 M^2 \Lambda_\chi^2} \nnu \\
&&\hspace{-0.5cm} 
+\ \left[2 \frac{m_\pi^2}{\Lambda_\chi^2} 
         - 6  \frac{M \left( m_u + m_d \right)}{\Lambda_\chi^2} \right]^2
+  O\left(m_\pi^6\right)
\eea
and
\bea
m_\pi^2\ =\ (m_\pi^0)^2 
&&\hspace{-0.5cm} 
-\ \frac{\left(m_u + m_d \right)^2}{\Lambda_\chi^2} B_0^2
+ 6 \frac{M \left(m_u + m_d \right)^2}{\Lambda_\chi^2} B_0  
- \left(m_u + m_d \right)^2 \nnu \\
&&\hspace{-0.5cm} 
+\ \frac{\left(m_u + m_d \right)^3}{M\ \Lambda_\chi^2} B_0^2 
- 6 \frac{m_u m_d \left(m_u + m_d \right)}{\Lambda_\chi^2} B_0
+ \ O\left(m_\pi^6,m_q^3\right)
\label{mch}
\eea
with 
\beq
B_0 \equiv - \frac{\langle \bar{q} q \rangle}{f^2} =
\frac{m_\pi^2}{\left(m_u + m_d \right)}.
\eeq
The decay
constants $f_\pi$ and $f_K$ are going to 
be necessary in determining $L_5$. The coupling $f_\pi$ computed  
in the chiral quark model is given at $ O(p^2,m_q)$ by
\beq
f_\pi = f \left[ 1 - \frac{f^2}{\langle \bar{q} q \rangle} 
\frac{m_\pi^2}{2 M} 
\left( 1 - 6 \frac{M^2}{\Lambda_\chi^2} \right) \right] \, .
\label{fch}
\eeq
The results for $f_\pi$ and $m_\pi^2$ 
computed from the strong chiral lagrangian by including the
relevant $O(p^4)$ counterterms (but no chiral loops) are,
\beq
f_\pi = f \left[ 1 + 4 \frac{ m_\pi^2}{f^2} L_5 + 
\frac{8 m_K^2 + 4 m_\pi^2}{f^2} L_4
\right],
\label{fsch}
\eeq
and
\beq
m_\pi^2 = (m_\pi^0)^2 + \frac{8}{f^2} B_0^2 \left(m_u + m_d\right)^2 \left[
2 L_8 - L_5 \right].
\label{msch} 
\eeq
 
Comparing  \eq{fch} and \eq{fsch} implies 
\beq
L_5 = - \frac{f^4}{8 {\langle \bar{q} q \rangle }} \frac{1}{M} \left( 1 - 6
\frac{M^2}{\Lambda_\chi^2}\right),
\label{L5}
\eeq
which is also obtained by a direct calculation within the $\chi$QM
(in the rotated picture of \cite{bef0}).
Accordingly, at the same level of approximation, we find
\beq
L_4 = 0.
\label{L4}
\eeq
 From \eq{mch} and \eq{msch} we get at $O(m_q^2)$
\beq
2 L_8 - L_5 = \frac{f^2}{ 8 B_0^2} \left[ - \frac{B_0^2} {\Lambda_\chi^2}  
+ 6 B_0 \frac{M}{\Lambda_\chi^2} - 1 \right] \ . \label{a.14}
\eeq
Eq.~(\ref{a.14}) together with \eq{L5}, yields
\beq
L_8 = - \frac{N_c}{16 \pi^2} \frac{1}{24}  - \frac{f^4}
{16 {\langle \bar{q} q \rangle } M} \left( 1 
+ \frac{M f^2}{{\langle \bar{q} q \rangle }}\right).
\label{L8}
\eeq

The coupling $f_K$ and $m_K$ could be equally used in
determining $L_4$, $L_5$ and $L_8$  leading to the same result.

\clearpage
\section{Input Parameters} 
\begin{table}[h]
\begin{center}
\begin{small}
\begin{tabular}{|c|c|}
\hline
{\rm parameter} & {\rm value} \\
\hline
$\sin ^2 \theta_W(m_Z)$ & 0.23 \\
$m_Z$ & 91.187 GeV \\
$m_W$ & 80.33 GeV \\
$m_t^{\rm pole}$ & $175 \pm 6$ GeV \\
$m_b(m_b)$ & 4.4 GeV \\
$m_c(m_c)$ & 1.4 GeV \\
$m_s$ (1 GeV) & $178 \pm 18$ MeV \\
$m_u + m_d$ (1 GeV) & $12 \pm 2.5$ MeV \\
$\Lambda_{QCD}^{(4)}$ & $340 \pm 40$ MeV \\
\hline
$V_{ud}$ & 0.9753 \\
$V_{us}$ & $0.221$ \\
\hline
$M$ & $200\ ^{+5}_{-3}$ MeV \\
$\vev{\bar{q}q}$  &  $- (240\ ^{+30}_{-10} \: \mbox{MeV} )^3$ \\
$ \langle \alpha_s GG/\pi \rangle $ & $(334 \pm 4 \: \mbox{MeV} )^4 $ \\
\hline
$f$  &  86 $\pm$ 13  MeV \\
$f_\pi = f_{\pi^+}$  &  92.4  MeV \\
$f_K = f_{K^+}$ & 113 MeV \\
$m_\pi = (m_{\pi^+} + m_{\pi^0})/2 $ & 138 MeV \\
$m_K = m_{K^0}$ &  498 MeV \\
$m_\eta$ & 548 MeV \\
$\cos\delta_0$ & $0.8$ \\
$\cos\delta_2$ & $1.0$ \\
\hline
\end{tabular}
\end{small}
\end{center}
\caption{Table of the numerical values of the input parameters.}
\end{table}

%
\clearpage
\renewcommand{\baselinestretch}{1}

\end{document}